\documentclass[10pt]{IEEEtran}
\usepackage{graphicx}
\usepackage{color}
\usepackage{dsfont}
\usepackage{amsmath}
\usepackage{cite}
\usepackage{float}
\usepackage{stfloats}
\usepackage{bbm}

\newcommand{\vb}[1]{\mathbf{#1}}
\newcommand{\vbhat}[1]{\vb{\hat #1}}
\newcommand{\numeq}[2]{\begin{equation} #2 \label{#1} \end{equation}}
\newcommand\sups[1]{^{\hbox{\scriptsize{#1}}}}
\newcommand\supt[1]{^{\hbox{\tiny{#1}}}}

\newcommand{\nn}{\nonumber \\}

\newcommand{\mc}{\mathcal}
\newcommand{\wt}{\widetilde}

\newcommand{\citeasnoun}[1]{Ref.~\citen{#1}}

\newcommand{\red}[1]{\textcolor{black}{#1}}

\begin{document}

\title {Generalized Taylor-Duffy Method for Efficient Evaluation
        of Galerkin Integrals in Boundary-Element Method Computations}
\author{M.~T.~Homer~Reid,
        \thanks{M. T. Homer Reid is with the Department of Mathematics, 
                Massachusetts Institute of Technology.}
        Jacob K. White, \textit{Fellow, IEEE},
        \thanks{J. K. White is with the Department of Electrical Engineering
                and Computer Science, Massachusetts Institute of Technology.}
        and Steven~G.~Johnson%
        \thanks{S. G. Johnson is with the Department of Mathematics,
                Massachusetts Institute of Technology.}
       }

\maketitle

\begin{abstract}
We present a generic technique, automated by computer-algebra 
systems and available as open-source software~\cite{TaylorDuffyCode}, 
for efficient numerical evaluation of a large family of 
singular and nonsingular 4-dimensional integrals over 
triangle-product domains, such as those arising in the 
boundary-element method (BEM) of computational electromagnetism. 
Previously, practical implementation of BEM solvers often required 
the aggregation of multiple disparate integral-evaluation 
schemes~\cite{Taylor2003, Duffy1982,
Taskinen2003, Jarvenpaa2006, TongChew2007, Klees1996, Cai2002,
Khayat2005, Ismatullah2008, Graglia2008, Polimeridis2010, 
Polimeridis2013,
Andra1997, SauterSchwab2010, Erichsen1998}
in order to treat all of the distinct types of integrals 
needed for a given BEM formulation; in contrast, our technique 
allows many different types of integrals to be handled by the 
\emph{same} algorithm and the same code implementation. Our 
method is a significant generalization of the Taylor--Duffy 
approach~\cite{Taylor2003,Duffy1982}, which was originally 
presented for just a single type of integrand; in addition 
to generalizing this technique to a broad class of integrands, 
we also achieve a significant improvement in its efficiency 
by showing how the \emph{dimension} of the final numerical 
integral may often reduced by one. In particular, if $n$ 
is the number of common vertices between the two triangles, 
in many cases we can reduce the dimension of the integral 
from $4-n$ to $3-n$, obtaining a closed-form analytical 
result for $n=3$ (the common-triangle case).
\end{abstract}


\color{black}

\section{Introduction}
\label{IntroductionSection}

The application of boundary-element methods 
\{BEM~\cite{Harrington93, Chew2009}, also known as the 
method of moments (MOM)\} to surfaces discretized into 
triangular elements commonly requires evaluating 
four-dimensional integrals over triangle-product 
domains of the form
\numeq{OriginalIntegral}
{ \mathcal{I}=
  \int_{\mc T} \, d\vb x \, \int_{\mc T^\prime} \, d\vb x^\prime \,
  P\big(\vb x, \vb x^\prime\big) K\big(|\vb x-\vb x^\prime|\big)
}
where $P$ is a polynomial, $K(r)$ is a kernel function which
may be singular at $r=0$, and 
$\mc T, \mc T^\prime$ are flat triangles; we will here be 
concerned with the case in which $\mc T,\mc T^\prime$ have
one or more common vertices. 
Methods for efficient and accurate evaluation of such
integrals have been extensively researched; among the 
most popular strategies are 
singularity subtraction 
(SS)~\cite{Taskinen2003, Jarvenpaa2006, TongChew2007},
singularity cancellation 
(SC)~\cite{Klees1996, Cai2002, Khayat2005, 
Ismatullah2008, Graglia2008},
and fully-numerical 
schemes~\cite{Polimeridis2010, Polimeridis2013}.
(Strategies have also been proposed to handle
the \textit{near-singular} case in which 
$\mc T, \mc T^\prime$ have vertices which are 
nearly but not precisely coincident~\cite{Botha2013, 
Vipiana2013}; we do not address that case here.)
Particularly interesting among SC methods is the scheme proposed 
by Taylor~\cite{Taylor2003} following earlier ideas of 
Duffy~\cite{Duffy1982} 
(see also Refs.~[\citen{Andra1997, SauterSchwab2010, Erichsen1998}]);
we will refer  to the method of~\citeasnoun{Taylor2003} as the
the ``Taylor-Duffy method'' (TDM). This method considered the 
specific kernel $K\sups{Helmholtz}(r)=\frac{e^{ikr}}{4\pi r}$
and a specific linear polynomial $P\sups{linear}$ and reduced
the singular 4-dimensional integral (\ref{OriginalIntegral})
to a nonsingular $(4-n)$-dimensional integral
(where $n\in \{1,2,3\}$ is the number of vertices common
to $\mc T, \mc T^\prime$)
with a complicated integrand obtained by performing
various manipulations on $K\sups{Helmholtz}$ and $P\sups{linear}$.
The reduced integral is then evaluated numerically
by simple cubature methods.

Our first objective is to show that the TDM
may be generalized to handle a significantly 
broader class of integrand functions.
Whereas~\citeasnoun{Taylor2003} addressed the
\textit{specific} case of the Helmholtz kernel 
combined with constant or linear factors, the 
master formulas we present [equations (2) in
Section \ref{GeneralizedTaylorDuffySection}]
are nonsingular reduced-dimensional versions of
(\ref{OriginalIntegral}) that apply to 
a broad \textit{family} of kernels $K$
combined with \textit{arbitrary} polynomials $P$.
Our master formulas (2) involve new functions 
$\mc K$ and $\mc P$ derived from $K$ 
and $P$ in (\ref{OriginalIntegral}) by procedures,
discussed in the main text and Appendices,
that abstract and generalize the techniques 
of~\citeasnoun{Taylor2003}.

We next extend the TDM by showing that, for some 
kernels---notably including the ``$r-$power'' kernel 
$K(r)=r^p$ for integer $p$---the reduction of 
dimensionality effected by the TDM may be carried 
one dimension further, so that the original 4-dimensional 
integral is converted into a $(3-n)$-dimensional integral
[equations (5) in Section
\ref{TwiceIntegrableKernelSection}].
In particular, in the common-triangle case $(n=3)$, we
obtain a \textit{closed-form analytical solution} 
of the full 4-dimensional integral 
(\ref{OriginalIntegral}). This result encompasses and 
generalizes existing results~\cite{Caorsi1993, Eibert1995}
for closed-form evaluations of the 
four-dimensional integral for certain
special $P$ and $K$ functions.

A characteristic feature of many published strategies
for evaluating integrals of the form (\ref{OriginalIntegral})
is that they depend on specific choices of the
$P$ and $K$ functions, with (in particular) each
new type of kernel understood to necessitate new
computational strategies. In practical implementations
this can lead to cluttered codes, requiring multiple
distinct modules for evaluating the integrals 
needed for distinct BEM formulations.
The technique we 
propose here alleviates this difficulty. Indeed, as we
discuss in Section \ref{BEMFormulationsSection}, the
flexibility of our generalized TDM allows the \textit{same} 
basic code \{$\sim$1,500 lines of C++ 
(not including general-purpose utility libraries),
available for download as free open-source 
software~\cite{TaylorDuffyCode}\} to handle 
\textit{all} singular integrals arising in
several popular BEM formulations. Although separate
techniques for computing these integrals have been 
published before, the novelty of our approach is 
to attack many different integrals with the \textit{same}
algorithm and the \textit{same} code implementation.

Of course, the efficiency and generality of the TDM 
reduction do not come for free: the cost is that the
reduction \textit{process}---specifically, the procedure 
by which the original polynomial $P$ in 
(\ref{OriginalIntegral}) is converted into new 
polynomials $\mc P$ that enter the master formulas 
(2) and (5)---is tedious and error-prone if 
carried out by hand. To alleviate this difficulty, we 
have developed a computer-algebra technique for 
automating this conversion
(Section \ref{ComputerAlgebraSection}); our procedure
inputs the coefficients of $P$ and emits code for 
computing $\mc P$, which may be directly
incorporated into routines for numerical 
evaluation of the integrands
of the reduced integrals (2) or (5).

\begin{figure*}
 \vspace*{-1.5in}
  \hspace{-1.0in}
  \includegraphics{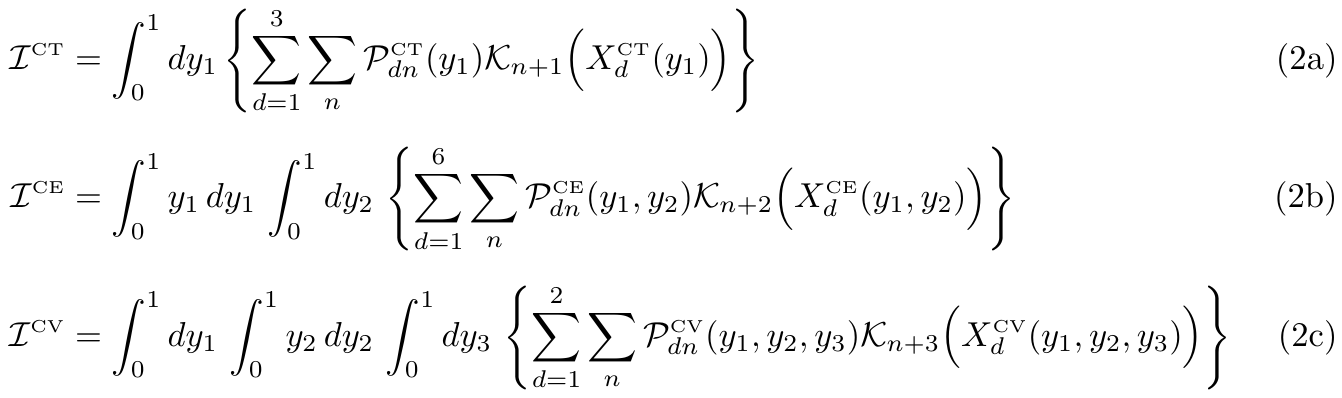}
 \vspace*{-7.5in}
\caption{The basic TDM for once-integrable kernels 
reduces the original integral (\ref{OriginalIntegral})
to a 1-, 2-, or 3-dimensional integral in the
\textbf{(a)} common-triangle, \textbf{(b)} common-edge, and 
\textbf{(c)} common-vertex cases respectively. 
(Note that this is an \textit{exact}, not approximate, 
reduction.)
The $\mc P$ and $\mc K$
functions in the integrands are derived from the $P$ and $K$ functions
in the original integral (\ref{OriginalIntegral}) by procedures
discussed in the main text.}
\medskip
\hrulefill
\end{figure*}
\setcounter{equation}{2}

The TDM reduces the 4-dimensional integral
(\ref{OriginalIntegral}) to a lower-dimensional
integral which is evaluated by numerical
cubature. How smooth is the reduced integrand,
and how rapidly does the cubature converge with
the number of integrand samples?
These questions are addressed in 
Section \ref{ExamplesSection}, where we 
plot integrands and convergence rates for
the reduced integrals resulting from applying
the generalized TDM to a number of practically
relevant cases of (\ref{OriginalIntegral}).
We show that---notwithstanding the presence of
singularities in the original integral or geometric
irregularities in the panel pair---the reduced 
integrand is typically a smooth, well-behaved 
function which succumbs readily to straightforward 
numerical cubature.

Although the TDM is an SC scheme, it has useful application
to SS schemes. In such methods one subtracts the first few terms
from the small-$r$ expansion of the Helmholtz kernels; the 
non-singular integral involving the subtracted kernel is 
evaluated by simple numerical cubature, but the integrals 
involving the singular terms must be evaluated by other
means. In Section \ref{CachingSection} we note that these
are just another type of singular integral of the form 
(\ref{OriginalIntegral}), whereupon they may
again be evaluated using the same generalized
TDM code---and, moreover, because the
kernel in these integrals is just the ``$r$-power''
kernel $K(r)=r^p$, the improved TDM reduction 
discussed in Section \ref{TwiceIntegrableKernelSection}
is available. We compare the efficiency of the unadorned
TDM to a combined TDM/SS method and note that the
latter is particularly effective for broadband
studies of the same structure at many
different frequencies.

Finally, in Section \ref{HighKSection} we 
note a curious property of the Helmholtz
kernel in the short-wavelength limit:
as $k\to\infty$: this kernel becomes
``twice-integrable'' (a notion discussed 
below), and the accelerated TDM scheme
of Section \ref{TwiceIntegrableKernelSection}
becomes available.
In particular, in the common-triangle case,
the full four-dimensional integral 
(\ref{OriginalIntegral}) with 
$K(r)=\frac{e^{ikr}}{4\pi r}$ and arbitrary
polynomial $P$ may be evaluated in closed
analytical form in this limit.

Our conclusions are presented in Section 
\ref{ConclusionsSection}, and a number of 
technical details are relegated to the 
Appendices. A free, open-source software 
implementation of the method presented
in this paper is available 
online~\cite{TaylorDuffyCode}.

\color{black}

\section{Master Formulas for the Generalized Taylor-Duffy Method}
\label{GeneralizedTaylorDuffySection}

\red{
\citeasnoun{Taylor2003} considered the integral 
(\ref{OriginalIntegral}) for \textit{specific} choices
of $K$ and $P$---namely, the Helmholtz 
kernel $K=K\sups{Helmholtz}$ and a certain linear polynomial
$P=P\sups{linear}$---and showed that the singular 
4-dimensional integral could be reduced to a 
nonsingular lower-dimensional integral with a complicated 
integrand obtained by performing various manipulations on 
$K\sups{Helmholtz}$ and $P\sups{linear}.$ 
The dimension of the reduced integral is $4-n$, 
where $n$ is the number of common vertices between 
$\mc T,\mc T^\prime$.
The objective of this section is to abstract and generalize 
the procedure of~\citeasnoun{Taylor2003} to handle integrals 
of the form (\ref{OriginalIntegral}) for \textit{general}
$K$ and $P$ functions.}

\red{
(More specifically---as discussed in~\citeasnoun{Taylor2003}---
the reduction proceeds by dividing the original four-dimensional 
region of integration into multiple subregions, introducing 
appropriate variable transformations for each subregion that 
allow one or more integrations to  
be performed analytically, then summing the results
for all subregions to obtain a single lower-dimensional
integral. A detailed review of this procedure, using
the generalized notation of this paper, may be found online
in the documentation for our open-source
code~\cite{TaylorDuffyCode}.)
}

\red{
The result is equation (2) at the top of the following page; 
here integrals (2)\{a,b,c\} are nonsingular integrals 
which evaluate to the same result as integral (1) for the 
common-triangle $(n=3)$, 
common-edge $(n=2)$, and 
common-vertex $(n=1)$ cases,
respectively. In particular, equation (2)a may be 
understood as a generalized version of equation (46)
in~\citeasnoun{Taylor2003}. 
}
In the integrals of equation (2),
\begin{itemize}

 \item \red{The domain of integration in each integral is 
            \textit{fixed} (it is the unit \{interval, square, cube\}
            for the \{CT, CE, CV\} case) independent of $P$, $K$,
            and the triangles $\{\mc T, \mc T^\prime\}$.
            Thus a numerical implementation need only furnish 
            cubature rules for these three fixed domains; there 
            is no need to construct custom-designed cubature rules 
            for particular triangles or integrand functions.
           }

 \item The $d$ index runs over subregions into which the
       original dimensional integration domain is 
       divided; there are $\{3, 6, 2\}$ subregions for the 
       \{CT, CE, CV\} cases.

 \item For each subregion $d$, the $X_d$ functions
       are ``reduced distance'' functions for that subregion.
       $X_d(\{y_i\})$ is the square root of a second-degree polynomial
       in the $y_i$ variables, whose coefficients depend on
       the geometrical parameters of the two triangles. 
       (Explicit expressions are given in Appendix 
       \ref{SubRegionAppendix}.)
       Note that the division into subregions, and the 
       $X_d$ functions, are independent of the specific 
       $P$ and $K$ functions in the original integrand.

 \item For each subregion $d$ and each integer $n$,
       the functions $\mathcal{P}_{dn}(y_i)$ are polynomials 
       derived from the original polynomial $P(\vb x, \vb x^\prime)$  
       in (\ref{OriginalIntegral}).
       For a given $P(\vb x, \vb x^\prime),$ the derived
       polynomials $\mathcal{P}_{dn}$ are only nonzero for 
       certain integers $n$; this defines the limits of the 
       $n$ summations in (2). 
       The procedure for obtaining $\mathcal{P}$ 
       from $P$ is discussed in 
       Section \ref{ComputerAlgebraSection} and 
       Appendix \ref{SubRegionAppendix}.

 \item For each integer $n$, the function $\mathcal{K}_n$ is obtained
       from the $K$ kernel as follows:
       \numeq{FirstIntegral}
       {\mathcal{K}_n(X) \equiv \int_0^1 w^n K(wX) \, dw. }
       For several kernels of interest, this integral may be 
       evaluated explicitly to obtain a closed-form expression
       for $\mathcal{K}_n$. We will refer to such kernels
       as \textit{once-integrable}. 
       Appendix \ref{FirstSecondIntegralAppendix} tabulates 
       the $\mathcal{K}_n$ functions for several once-integrable
       kernel functions. (In the following section
       we will introduce the further notion of
       \textit{twice-integrability.})
\end{itemize}
The key advantage of the TDM is that equation (\ref{FirstIntegral})
isolates the integrable singularities in (\ref{OriginalIntegral}) 
into a one-dimensional integral which may be performed analytically. 
This not only reduces the dimension of the original integral 
(\ref{OriginalIntegral}), but also neutralizes its singularities, 
leaving behind a smooth integrand. (In the CT and CE cases, the 
dimension of the integral may be reduced further.)  The remaining 
integrals (2), though complicated, are amenable to efficient evaluation 
by numerical cubature.

\red{ The \textit{extent} to which the singularities 
      of $K$ are regulated depends on the polynomial $P$.
      More specifically, if the original kernel $K(r)$
      diverges for small $r$ like $r^{-L}$, then
      equation (\ref{FirstIntegral}) will be finite
      only for $n\ge L$. As noted above, the range 
      $[n\sups{min},n\sups{max}]$ of integers $n$ for 
      which we need to compute (\ref{FirstIntegral}) 
      depends on the number of common vertices 
      and on the polynomial $P$ (in particular, 
      $n\sups{min}$ is greater for polynomials that 
      vanish at $\vb x=\vb x^\prime$).
      For example, in the common-triangle case
      with polynomial $P\equiv 1$ one finds
      $n\sups{min}=1$; thus in this case we can 
      integrate kernels $K(r)$ that behave for 
      small $r$ like $1/r$, but not kernels
      that diverge like $1/r^2$ or faster.
      On the other hand, in the common-edge case
      with $P=P\sups{MFIE}$
      (Section \ref{BEMFormulationsSection})
      one finds $n\sups{min}=3$, allowing treatment
      of kernels with $1/r^3$ singularities
      such as 
      $K\sups{MFIE}(r)=(ikr-1)\frac{e^{ikr}}{4\pi r^3}$.
    }

\section{Improved TDM Formulas for Twice-Integrable Kernels}
\label{TwiceIntegrableKernelSection}

The TDM reduces the original 4-dimensional integral 
(\ref{OriginalIntegral}) to the $(4-n)$-dimensional integral
(2) where $n\in\{1,2,3\}$ is the number 
of common vertices. In this section we show that, for certain 
kernel functions, it is possible to go further; 
when the kernel is \textit{twice-integrable}, in a sense
defined below, the original 4-dimensional integral 
is reduced to a $(3-n)$-dimensional integral. In particular,
for the case $n=3$, the full 4-dimensional integral
may be evaluated explicitly to yield a \textit{closed-form 
expression} requiring no numerical integrations.

The master TDM formulas for twice-integrable kernels
are equations (5) at the top of the following page,
and their derivation is discussed below.

\subsection*{Twice-Integrable Kernels}

Above we referred to a kernel function $K(r)$ as 
\textit{once integrable} if it is possible to evaluate the 
integral (\ref{FirstIntegral}) in closed form.
For such kernels, we now introduce a further qualification: 
we refer to $K(r)$ as \textit{twice-integrable}
if it is possible to obtain closed-form expressions for the 
following two integrals involving the $\mathcal{K}$ function
defined by (\ref{FirstIntegral}):
\begin{subequations}
\begin{align}
  \mathcal{J}_n(\alpha,\beta,\gamma) 
&\equiv 
  \int_0^1 \mathcal{K}_n\Big( \alpha\sqrt{ (y+\beta)^2 + \gamma^2} \Big) \, dy
\\
  \mathcal{L}_n(\alpha,\beta,\gamma) 
&\equiv 
  \int_0^1 y\, \mathcal{K}_n\Big( \alpha\sqrt{ (y+\beta)^2 + \gamma^2} \Big) \, dy.
\end{align}
\label{SecondIntegrals}
\end{subequations}

\noindent In particular, the kernel $K(r)=r^p$ is 
twice-integrable for arbitrary integer powers $p$;
moreover, in Section \ref{HighKSection}
we show that the Helmholtz kernels
become twice-integrable in the limit $\text{Im } k\to \infty.$
(Expressions for $\mathcal{J}$ and $\mathcal{L}$
in all these cases are collected in Appendix \ref{FirstSecondIntegralAppendix}.)

\begin{figure*}
 \vspace*{-1.5in}
  \hspace{-1.0in}
  \includegraphics{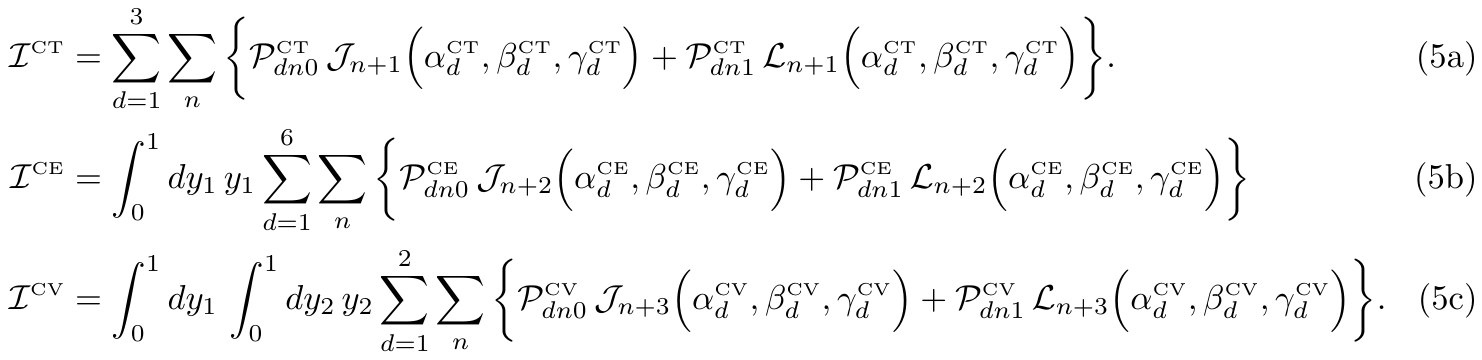}
 \vspace*{-7.5in}
\caption{The TDM for twice-integrable kernels 
reduces the original integral (\ref{OriginalIntegral})
to a 0-, 1-, or 2-dimensional integral in the
\textbf{(a)} common-triangle, \textbf{(b)} common-edge, and 
\textbf{(c)} common-vertex cases respectively.}

\medskip

\hrulefill
\end{figure*}

\subsection*{The TDM For Twice-Integrable Kernels}

For twice-integrable kernels, the formulas (2)
may be further simplified by analytically evaluating the
innermost integral in each case.
Thus, for the \{CT, CE, CV\} case, we analytically
perform the $\{y_1, y_2, y_3\}$ integration.

We will consider here the case in which the $\mc P$ polynomials 
are of degree not greater than 1 in the innermost integration 
variable.  (This condition is satisfied, in particular, for all
but one of the eight distinct forms of the $P$ polynomials considered in
Section \ref{BEMFormulationsSection}.) More general cases could be
handled by extending the methods of this section.

\subsubsection*{The Common-Triangle Case}

Given the above assumption on the degree of the $\mc P$ polynomials,
we can write, in the common-triangle case,
\setcounter{equation}{5}
\numeq{PCTExpansion}
{ \mc P\supt{CT}_{dn}(y) = 
 \mc P\supt{CT}_{dn0} + y \mc P\supt{CT}_{dn1},
}
where $\mc P\supt{CT}_{dn0}$ and $\mc P\supt{CT}_{dn1}$ are
just the constant and linear coefficients in the polynomial 
$\mc P\supt{CT}_{dn}(y)$. 
The Taylor-Duffy formula for the common-triangle case,
equation (2a), then becomes 
\begin{align}
 \mathcal{I}\supt{CT}
=
  \sum_{d=1}^3 \sum_{n}
  \bigg\{ 
  &\mathcal{P}\supt{CT}_{dn0} 
   \int_0^1 dy \, \mathcal{K}_{n+1}\Big( X\supt{CT}_d(y) \Big)
\nn
  &\quad+\mathcal{P}\supt{CT}_{dn1} \,
   \int_0^1 dy \, y\,\mathcal{K}_{n+1}\Big( X\supt{CT}_d(y) \Big)
  \bigg\}.
 \label{TDMFormulas20}
\end{align}
If we now write the reduced-distance function $X\supt{CT}_d(x)$ 
in the form (Appendix \ref{SubRegionAppendix})
\numeq{XCTReduced}
{X\supt{CT}_d(y) \equiv 
 \alpha\supt{CT}_d\sqrt{(y+\beta\supt{CT}_d)^2 + (\gamma\supt{CT}_d)^2 }
}
(where $\alpha_d, \beta_d$, and $\gamma_d$ are functions of
the geometrical parameters such as triangle side lengths and
areas)
then we can immediately use equations (\ref{SecondIntegrals}) to
evaluate the $y$ integrals in (\ref{TDMFormulas20}),
obtaining an exact closed-form expression for the full 4-dimensional
integral (\ref{OriginalIntegral}) in the common-triangle case.
This is equation (5a).

We emphasize again that (5a) involves no
further integrations, but is a
\textit{closed-form expression for the full 4-dimensional integral}
in (\ref{OriginalIntegral}). 
Closed-form expressions for certain special cases of 
4-dimensional triangle-product integrals in BEM schemes have 
appeared in the literature before~\cite{Caorsi1993, Eibert1995},
but we believe equation (5a) to be the most
general result available to date.

\subsubsection*{The Common-Edge and Common-Vertex Cases}

We now proceed in exactly analogous fashion for the common-edge and 
common-vertex cases. We will show that, for
twice-integrable kernels, the 2-dimensional and 3-dimensional 
integrals obtained via the usual TDM 
[equations (2b, 2c)] 
may be reduced to 1-dimensional and 2-dimensional integrals,
respectively.

Because the 
$\mc P\supt{CE}$ and $\mc P\supt{CV}$ polynomials are (by assumption)
not more than linear in the variables $y_2$ and $y_3$, respectively, 
we can write, in analogy to (\ref{PCTExpansion}),
\begin{subequations}
\begin{align}
  \mc P\supt{CE}_{dn}(y_1, y_2)
&=
  \mc P\supt{CE}_{dn0} + y_2 \mc P\supt{CE}_{dn1}
\\[5pt]
  \mc P\supt{CV}_{dn}(y_1, y_2, y_3)
&=
  \mc P\supt{CV}_{dn0} + y_3 \mc P\supt{CV}_{dn1}
\end{align}
\label{PCEPCV}
\end{subequations}
where the ${\mc P}\supt{CE}_{dni}$ coefficients 
depend on $y_1$ in addition to the geometric 
parameters, while the ${\mc P}\supt{CV}_{dni}$ 
coefficients depend on $y_1$ and $y_2$
in addition to the geometric parameters.

Similarly, in analogy to (\ref{XCTReduced}), we write
\begin{subequations}
\begin{align}
 X\supt{CE}_d(y_1, y_2) 
   \equiv 
   \alpha\supt{CE}_d \sqrt{(y_2+\beta\supt{CE}_d)^2 + (\gamma\supt{CE}_d)^2 }
\\
X\supt{CV}_d(y_1, y_2, y_3) 
   \equiv 
   \alpha\supt{CV}_d \sqrt{ (y_3+\beta\supt{CV}_d)^2 + (\gamma\supt{CV}_d)^2 }.
\end{align}
\label{XCEXCV}
\end{subequations}
where $\{\alpha, \beta, \gamma\}\supt{CE}_{d}$ depend on 
$y_1$ in addition to the geometric parameters, while 
$\{\alpha, \beta, \gamma\}\supt{CV}_{d}$ depend on 
$y_1$ and $y_2$ in addition to the geometric parameters.

Inserting (\ref{PCEPCV}) and (\ref{XCEXCV}) into
(2b) and (2c) and evaluating the $y_2$ and
$y_3$ integrals using (\ref{SecondIntegrals}), 
the original 4-dimensional integral (\ref{OriginalIntegral})
is then reduced to a 
1-dimensional integral [equation (5b)] 
or a 
2-dimensional integral [equation (5c)].

Thus, for twice-integrable kernels, the dimension of the
numerical cubature needed to evaluate the original integral 
(\ref{OriginalIntegral}) is reduced by 1 compared to 
the case of once-integrable kernels.

\subsection*{Summary of Master TDM Formulas}

For once-integrable kernel functions,
the generalized TDM reduces the original
4-dimensional integral (\ref{OriginalIntegral}) to
a $(4-n)$-dimensional integral, equation 
(2), where $n$ is the number
of common vertices between the triangles.

For twice-integrable kernel functions, 
the generalized TDM reduces
the original four-dimensional integral
(\ref{OriginalIntegral}) to 
a $(3-n)$-dimensional integral, equation (5). 
In particular, in the common-triangle case 
$n=3$ we obtain a \textit{closed-form expression} 
requiring no numerical integrations.

In addition to reducing the dimension of the 
integral, the TDM also performs the service
of neutralizing singularities that may be 
present in the original four-dimensional 
integral, ensuring that the resulting
integrals (2) or (5)
are amenable to efficient evaluation by 
numerical cubature.

\section{From $P$ to $\mathcal{P}$: Computer Algebra Techniques}
\label{ComputerAlgebraSection}

The integrands of the Taylor-Duffy integrals 
(2) and (5) refer to
polynomials $\mathcal{P}$ derived from the original polynomial
$P$ appearing in the original integral (\ref{OriginalIntegral}). The 
procedure for obtaining $\mathcal{P}$ from $P$, summarized
in equations 
(\ref{CommonTriangleHToP}),
(\ref{CommonEdgeHToP}),
and 
(\ref{CommonVertexHToP}),
is straightforward but tedious and error-prone if carried 
out by hand. 
For example, to derive the polynomials 
$\mathcal{P}_{dn}\supt{CT}$ in the common-triangle 
formulas (2a) and (5a),
we must
\textbf{(a)} define, for each subregion $d=1,2,3$, 
a new function $H_d(u_1, u_2)$ by
evaluating a certain definite integral involving 
the $P$ polynomial, 
\textbf{(b)} evaluate the function $H$
at certain $w$-dependent arguments to obtain
a polynomial in $w$, and then 
\textbf{(c)} identify 
the coefficients of $w^n$ in this polynomial
as the $\mathcal{P}_{dn}\supt{CT}$ functions we seek.
Moreover, we must repeat this procedure for each of the 
three subregions that enter the common-triangle case, and
for the common-edge case we have \textit{six} subregions. 
Clearly the process of reducing (\ref{OriginalIntegral})
to (2) or (5) is too complex a task to entrust 
to pencil-and-paper calculation.

However, the manipulations are 
ideally suited to evaluation by 
\textit{computer-algebra} systems.
For example, Figure 3
presents {\sc mathematica} code that executes the 
procedure described above for deriving the 
$\mathcal{P}_{dn}\supt{CT}$ polynomials
for one choice of $P(\vb x, \vb x^\prime)$ 
function (specifically, the polynomial named
$P\sups{EFIE1}$ in Section \ref{BEMFormulationsSection}).
Running this script yields the emission of 
machine-generated code for computing the $\mc P$
polynomials, and this code may be directly 
incorporated into a routine for computing the
integrands of (2) or (5). \red{(This and
other computer-algebra codes for automating
the procedures of this paper may be found
together with our online open-source
code distribution~\cite{TaylorDuffyCode}.)}

%
%

\begin{figure}[t]
\begin{verbatim}

(*****************************************)
(* P polynomial for the case             *)
(*  P(x,xp) = (x-Q) \cdot (xp - QP)      *)
(*****************************************)
P[Xi1_, Xi2_, Eta1_, Eta2_] :=           \
  Xi1*Eta1*A*A + Xi1*Eta2*AdB + Xi1*AdDP \
+ Xi2*Eta1*AdB + Xi2*Eta2*B*B + Xi2*BdDP \
+     Eta1*AdD +     Eta2*BdD +     DdDP;

(*****************************************)
(* region-dependent integration limits   *)
(* and u-functions, Eqs (21) and (23)    *)
(*****************************************)
u1[d_, y_]:=Switch[d, 1, 1, 2, y,   3, y];
u2[d_, y_]:=Switch[d, 1, y, 2, y-1, 3, 1];

Xi1Lower[d_, u1_, u2_]                   \
 := Switch[ d, 1, 0, 2, -u2, 3, u2-u1];
Xi1Upper[d_, u1_, u2_]                   \ 
 := 1-u1;
Xi2Lower[d_, u1_, u2_, Xi1_]             \
 := Switch[ d, 1, 0, 2, -u2, 3, 0];  
Xi2Upper[d_, u1_, u2_, Xi1_]             \
  := Switch[ d, 1, Xi1, 2, Xi1,          \
                3, Xi1-(u1-u2)];

(*****************************************)
(* big H function, equation (22) *********)
(*****************************************)
H[u1_, u2_] :=                   \
 Integrate[                      \
  Integrate[                     \
    P[Xi1, Xi2, u1+Xi1, u2+Xi2]  \
  + P[u1+Xi1,u2+Xi2, Xi1, Xi2 ], \
   {Xi2Lower, 0, Xi2Upper }],    \
  {Xi1Lower, 0, Xi1Upper}];

(*****************************************)
(* \mathcal{P}_{dn} functions, eq. (24)  *)
(*****************************************)
wSeries = 
 Series[ H[ w*u1[x], w*u2[x] ], {w,0,10} ];
P[n_,x_] := SeriesCoefficient[ wSeries, n];
\end{verbatim}
 \label{Listing1} 
 \caption{{\sc mathematica} code snippet that derives 
          the $\mathcal{P}_{dn}\supt{CT}$ polynomials
          for a given $P(\vb x, \vb x^\prime)$ polynomial.}
\end{figure}

%
%

\newcommand{\mm}{\!\!-\!\!}
\newcommand{\ttimes}{\!\times\!}
\newcommand{\ccdot}{\!\cdot\!}

\section{\red{Applications to BEM Formulations using Triangle-Supported
              Basis Functions}}
\label{BEMFormulationsSection} 

\red{
As noted in the Introduction, a key strength of our
proposal is that the flexibility of our generalized TDM 
allows the \textit{same} code to evaluate singular integrals 
for many \textit{different} BEM formulations. Indeed,
our implementation~\cite{TaylorDuffyCode} ($\sim$1,500
lines of C++, not including utility libraries) suffices
to handle \textit{all} singular integrals arising in 
several popular BEM formulations:
\textbf{(a)} electrostatics with triangle-pulse
basis functions~\cite{Sarkar1984},
\textbf{(b)} the electric-field integral equation (EFIE)
with RWG basis functions~\cite{RWG1982},
\textbf{(c)} the magnetic-field integral equation 
(MFIE)~\cite{Rius2001} or PMCHWT~\cite{Medgyesi1994}
formulations with RWG basis functions,
\textbf{(d)} the n-M\"uller formulation with
RWG basis functions~\cite{Taskinen2005}.
[Moreover, in cases \textbf{(b)}--\textbf{(d)}, the
same code implementation evaluates not only the full-wave
integrals but also the individual contributions to those
integrals needed for singularity-subtraction schemes;
this is discussed in Section \ref{CachingSection}.]
}

\red{Of course, it is not a new result that singular integrals
of the form (\ref{OriginalIntegral}) may be reduced
to nonsingular lower-dimensional integrals; indeed,
the reduction of case \textbf{(b)} was the subject
of the original TDM paper~\cite{Taylor2003}, while
numerous works have pursued other specialized approaches
for the other cases~\cite{Taskinen2003, Jarvenpaa2006, 
TongChew2007, Klees1996, Cai2002, Khayat2005, 
Ismatullah2008, Graglia2008, Polimeridis2010, 
Polimeridis2013, Andra1997, SauterSchwab2010, 
Erichsen1998}. 
The novelty of our contribution here is the observation
that this proliferation of specialized approaches
is in fact unnecessary; instead, \textit{all} integrals 
arising in the four formulations above may be written 
in the form of equation (\ref{OriginalIntegral}),
whereupon they may be reduced by the
same generalized procedure to the
same general reduced form
[equations (2) or (7)] and then evaluated
using the same generalized code. This is an
advantage over the multiple distinct codes that 
would be required to implement each of the separate
methods of the references cited above.
}

\red{For reference, in this section we note the 
particular forms of the $P$ and $K$ functions
needed to write Galerkin integrals for
various BEM formulations in the form of 
equation (\ref{OriginalIntegral}).
These $P$ and $K$ functions may then be transformed, 
via methods discussed elsewhere in this paper,
into the $\mc P$ and $\mc K$ functions in the 
integrands of (2) and (7). Source-code implementations
of these functions 
for all formulations mentioned above may be found
in our open-source code~\cite{TaylorDuffyCode}.
}

\subsection*{Electrostatics with triangle-pulse functions}

For electrostatic BEM formulations using ``triangle-pulse'' basis 
functions representing constant charge densities on flat triangular
panels, we require the average over triangle $\mc T$ of the
potential and/or normal electric field due to a constant charge
density on $\mc T^\prime$. These are
\begin{subequations}
\begin{align}
\mathcal{I}\sups{ES1}
&=
  \int_{\mc T} \, \int_{\mc T^\prime} \,
  \frac{1}{4\pi|\vb x - \vb x^\prime|}\,
  \, d\vb x \, d\vb x^\prime
\\[5pt]
\mathcal{I}\sups{ES2}
&=
  \int_{\mc T} \, \int_{\mc T^\prime} \,
  \frac{\vbhat{n}\cdot(\vb x-\vb x^\prime)}
       {4\pi|\vb x - \vb x^\prime|^3}
        \,d\vb x \, d\vb x^\prime.
\end{align}
\label{ESIntegrals}
\end{subequations}
Equations (\ref{ESIntegrals}a) and (\ref{ESIntegrals}b) are of the form 
(\ref{OriginalIntegral}) with 
$$\begin{array}{lclclcl}
  \displaystyle{
  P\sups{ES1}(\vb x, \vb x^\prime)
               }
&=&
  \displaystyle{
  1,
               }
&\quad&
  \displaystyle{
  K\sups{ES1}(r)
               }
&=&
  \displaystyle{
 \frac{1}{4\pi r},
               }
\\[8pt]
  \displaystyle{
  P\sups{ES2}(\vb x, \vb x^\prime)
               }
&=&
  \displaystyle{
  \vbhat{n} \cdot (\vb x-\vb x^\prime),
               }
&\quad&
  \displaystyle{
   K\sups{ES2}(r)
               }
&=&
  \displaystyle{
   \frac{1}{4\pi r^3}.
               }
\end{array}$$

\subsection*{EFIE with RWG functions}

For the EFIE formulation of full-wave electromagnetism
with RWG source and test functions~\cite{RWG1982}, 
we require the electric field due to an
RWG distribution on $\mc T^\prime$ averaged over an RWG
distribution on $\mc T$. This involves the integrals
\begin{subequations}
\begin{align}
I\sups{EFIE1}
&=
 \frac{1}{4AA^\prime}
 \int_{\mc T} \, \int_{\mc T^\prime}  \,
 (\vb x-\vb Q)\cdot (\vb x^\prime-\vb Q^\prime)
 \frac{e^{ik|\vb x-\vb x^\prime|}}{4\pi|\vb x-\vb x^\prime|}
 \, d\vb x \, d\vb x^\prime
\\[10pt]
 I\sups{EFIE2}
&=
 \frac{1}{4AA^\prime}
 \int_{\mc T} \, \int_{\mc T^\prime}  \,
 \frac{e^{ik|\vb x-\vb x^\prime|}}{4\pi|\vb x-\vb x^\prime|}
 \, d\vb x \, d\vb x^\prime
\end{align}
\label{EFIEIntegrals}
\end{subequations}
where $A,A^\prime$ are the areas of $\mc T,\mc T^\prime$
and $\vb Q,\vb Q^\prime$ (the source/sink vertices of the 
RWG basis functions) are vertices in $\mc T, \mc T^\prime$.

Equations (\ref{EFIEIntegrals}) are of the form
(\ref{OriginalIntegral}) with
$$\begin{array}{lcllcl}
 \displaystyle{
 P\sups{EFIE1}(\vb x, \vb x^\prime)
              }
 &\!\!\!\!=\!\!\!\!&
 \displaystyle{
 \frac{(\vb x\!\!-\!\!\vb Q) \cdot (\vb x^\prime \!\!-\!\! \vb Q^\prime)}
         {4AA^\prime},
              }
 &
 \displaystyle{
   K\sups{EFIE1}(r)
              }
 &\!\!=\!\!&
 \displaystyle{
    \frac{e^{ikr}}{4\pi r}.
              }
\\[8pt]
 \displaystyle{
 P\sups{EFIE2}(\vb x, \vb x^\prime)
              }
 &\!\!\!\!=\!\!\!\!&
 \displaystyle{
 \frac{1}{AA^\prime},
              }
 &
 \displaystyle{
   K\sups{EFIE2}(r)
              }
 &\!\!=\!\!&
 \displaystyle{
    \frac{e^{ikr}}{4\pi r}
              }
\end{array}$$
[We will use the labels $K\sups{EFIE}$ 
and $K\sups{Helmholtz}$ interchangeably to 
denote the kernel $K(r)=\frac{e^{ikr}}{4\pi r}.$]

\subsection*{MFIE / PMCHWT with RWG functions}

For the MFIE formulation of full-wave electromagnetism
with RWG source and test functions~\cite{Rius2001}, 
we require the magnetic field due to an RWG distribution 
on $\mc T^\prime$ averaged over an RWG distribution on 
$\mc T$. This involves the integral
\numeq{MFIEIntegral}
{ I\sups{MFIE}
  =\!
  \frac{1}{4AA^\prime} 
  \int_{\mc T} \, \int_{\mc T^\prime}  \,
  (\vb x \!-\! \vb Q) \cdot \nabla \ttimes
  \left\{ (\vb x^\prime\!\!-\!\!\vb Q^\prime) 
          \frac{e^{ik|\vb x \!-\! \vb x^\prime|}}
               {4\pi|\vb x \mm \vb x^\prime|}
  \right\}
  \, d\vb x \, d\vb x^\prime.
}
With some rearrangement, equation (\ref{MFIEIntegral}) may
be written in the form (\ref{OriginalIntegral}) with
\begin{equation}
  P\sups{MFIE}(\vb x, \vb x^\prime)=
  \frac{  (\vb x- \vb x^\prime) \cdot (\vb Q\times \vb Q^\prime)
         +(\vb x \times \vb x^\prime) \cdot (\vb Q-\vb Q^\prime)
       }{4AA^\prime},
\end{equation}
\begin{equation}
   K\sups{MFIE}(r)
   =(ikr-1)\frac{e^{ikr}}{4\pi r^3}.
\end{equation}
With the EFIE and MFIE integrals, equations
(\ref{EFIEIntegrals}) and (\ref{MFIEIntegral}),
we also have everything needed to implement the PMCHWT formulation 
of full-wave electromagnetism with RWG source and test 
functions~\cite{Medgyesi1994}.

\subsection*{N-M\"uller with $\vbhat{n}\times$RWG/RWG functions}

For the N-M\"uller formulation with RWG basis functions
and $\vbhat{n}\times$RWG testing functions~\cite{Taskinen2005},
we require the electric and magnetic fields due to an 
RWG distribution on $\mc T^\prime$ averaged over an 
$\vbhat{n}\times$RWG distribution on $\mc T$; here $\vbhat{n}$
denotes the surface normal to $\mc T$.
These quantities involve the following integrals.
(We have here introduced the shorthand notation
${\wt \vb V } \equiv \vbhat{n}\times \vb V.$)
\begin{subequations}
\begin{align}
 I\sups{NM\"uller1}
&
\\
&\hspace{-0.3in}=
    \frac{1}{4AA^\prime} 
    \int_{\mc T} \, \int_{\mc T^\prime} \, 
    \big(\wt{\vb x}-\wt{\vb Q}\big)\cdot 
    \big(\vb x^\prime-\vb Q^\prime\big)
    \frac{e^{ik|\vb x-\vb x^\prime|}}
         {4\pi|\vb x-\vb x^\prime|}
    d\vb x \, d\vb x^\prime
\nonumber\\[15pt]
 I\sups{NM\"uller2}
&
\\
&\hspace{-0.4in}=
  \frac{1}{4AA^\prime} 
  \int_{\mc T} \! \int_{\mc T^\prime} \!
  (\wt{\vb x} \!\!-\!\! \wt{\vb Q}) \! \cdot \! \nabla
  \left\{ \big[\nabla^\prime \! \cdot \! 
           (\vb x^\prime\!\!-\!\!\vb Q^\prime)\big]
          \frac{e^{ik|\vb x \!-\! \vb x^\prime|}}
               {4\pi|\vb x \mm \vb x^\prime|}
  \right\}\!
  d\vb x d\vb x^\prime
\nonumber\\[8pt]
&\hspace{-0.3in}
 =\frac{2}{4AA^\prime}
    \int_{\mc T} \! \int_{\mc T^\prime} \! 
    \Big[ \big(\wt{\vb x}-\wt{\vb Q}\big) \cdot 
              (\vb x - \vb x^\prime)
    \Big]
    (ikr\mm1)\frac{e^{ikr}}{4\pi r^3}
  d\vb x d\vb x^\prime
\nonumber\\[15pt]
 I\sups{NM\"uller3}
&
\\
&\hspace{-0.3in}=
  \frac{1}{4AA^\prime} 
  \int_{\mc T} \, \int_{\mc T^\prime}  \,
  (\wt{\vb x} \mm \wt{\vb Q}) \ccdot \nabla \ttimes
  \left\{ (\vb x^\prime \mm \vb Q^\prime) 
          \frac{e^{ik|\vb x \!-\! \vb x^\prime|}}
               {4\pi|\vb x \mm \vb x^\prime|}
  \right\}
  \, d\vb x \, d\vb x^\prime
\nonumber
\end{align}
\label{NMullerIntegrals}
\end{subequations}

\noindent
Equations (\ref{NMullerIntegrals}) are of the form 
(\ref{OriginalIntegral}) with 
\begin{align*}
 P\sups{NM\"uller1}(\vb x, \vb x^\prime)
 &=\frac{ (\wt{\vb x} - \wt{\vb Q}) \cdot 
          (\vb x^\prime - \vb Q^\prime)}
        {4AA^\prime}, 
\\[3pt]
 P\sups{NM\"uller2}(\vb x, \vb x^\prime)
 &=\frac{ (\wt{\vb x}    - \wt{\vb Q}) \cdot 
           (\vb x - \vb x^\prime)}{2AA^\prime},
\\[3pt]
 P\sups{NM\"uller3}(\vb x, \vb x^\prime)
 &= \big(\wt{\vb x} - \wt{\vb Q}\big) \cdot 
      \Big[ \big(\vb x - \vb x^\prime\big)\times
            \big(\vb x^\prime - \vb Q^\prime\big)
      \Big],
\\[3pt]
 K\sups{NM\"uller1}(r)
 &=\frac{e^{ikr}}{4\pi r},
\\
 K\sups{NM\"uller2}(r)
 &=
 K\sups{NM\"uller3}(r)
 =(ikr-1)\frac{e^{ikr}}{4\pi r^3}.
\end{align*}

\section{Computational Examples}
\label{ExamplesSection}

In this section we consider a number of simple examples
to illustrate the practical efficacy of the generalized
TDM. For generic instances of the common-triangle,
common-edge, and common-vertex cases, we study the 
convergence vs. number of cubature points in the
numerical evaluation of integrals (2) or (5), and 
we plot the 1D or 2D integrands in various cases to
lend intuition for the function that is being integrated.
 
\subsection{Common-triangle examples}

\begin{figure}
\begin{center}
\resizebox{0.5\textwidth}{!}{\includegraphics{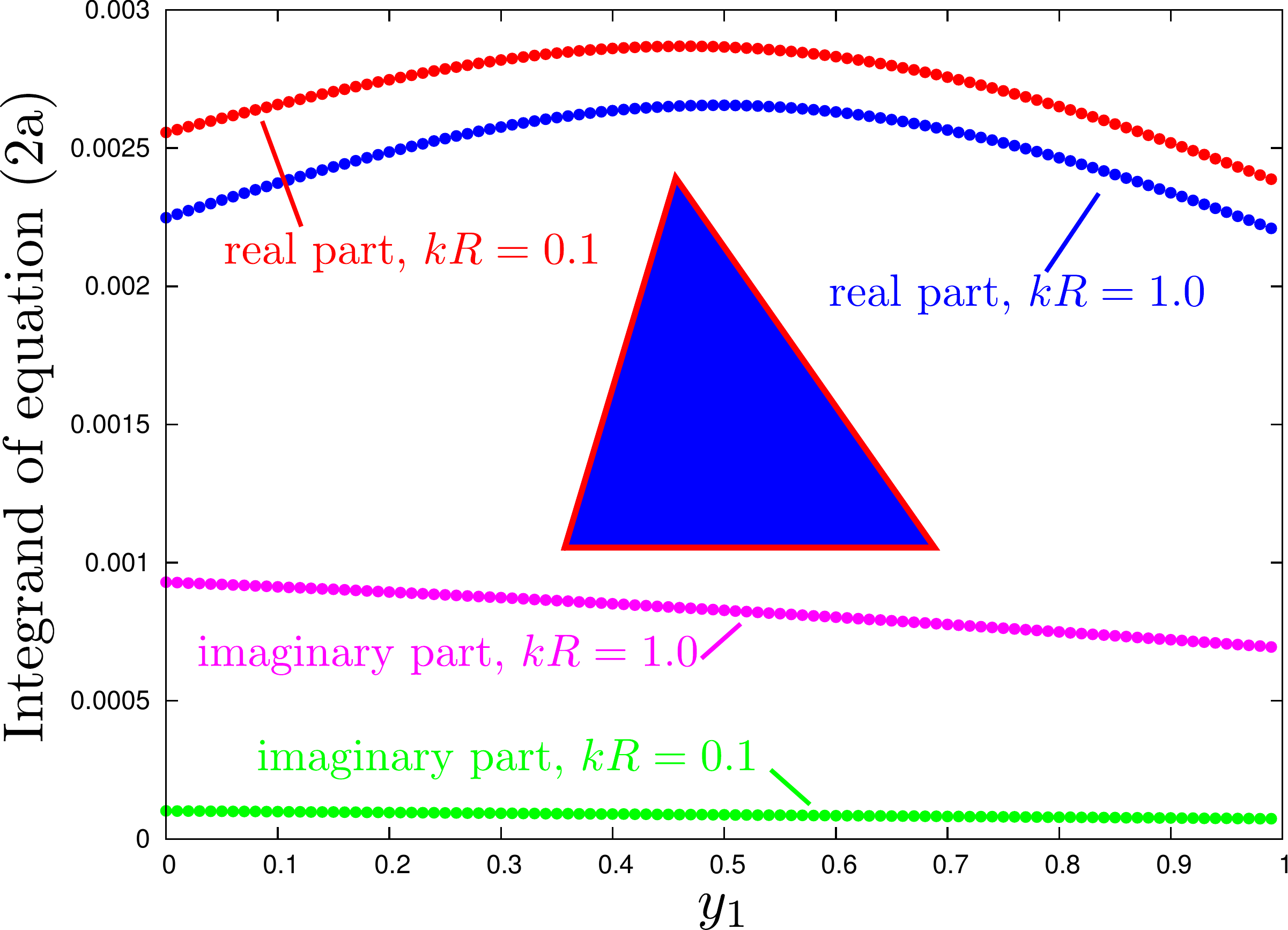}}
\caption{TDM integrand in the common-triangle case for a 
         once-integrable kernel. Plotted is the integrand 
         of equation (2a) with polynomial
         $P=P\supt{EFIE1}$ and kernel $K=K\supt{EFIE}$
         (Section \ref{BEMFormulationsSection}) for two different
         values of the wavenumber $k$. The triangle
         lies in the $xy$ plane with vertices at the points
         $(x,y)=(0,0), (0.1,0), (0.03,0.1).$ 
         The integrand is smooth and amenable to low-order 
         numerical quadrature.
        }
\label{CTExampleFigure}
\end{center}
\end{figure}
Figure \ref{CTExampleFigure} plots the integrand of equation (2a) for 
the choice of polynomial 
$P\sups{EFIE1}(\vb x, \vb x^\prime)
 \propto (\vb x-\vb Q)\cdot (\vb x-\vb Q^\prime)$
and kernel $K\sups{EFIE}(r)=\frac{e^{ikr}}{4\pi r}$,
a combination which arises in the EFIE formulation with RWG
functions (Section \ref{BEMFormulationsSection}).
The triangle (inset) lies in 
the $xy$ plane with vertices at the points 
$(x,y)=\{(0,0), (0.1,0), (0.03,0.1)\}$ with 
RWG source/sink vertices $\vb Q=\vb Q^\prime=(0,0).$
The wavenumber parameter $k$ in 
the Helmholtz kernel is chosen such that $kR=0.1$ or $kR=1.0$, where 
$R$ is the radius of the triangle (the maximal distance from centroid 
to any vertex). Whereas the integrand of the original integral 
(\ref{OriginalIntegral}) exhibits both singularities and sinusoidal 
oscillations over its 4-dimensional domain, the integrand of the 
TDM-reduced integral (2a) is nonsingular and slowly varying
and will clearly succumb readily to numerical quadrature; indeed, 
for both values of $k$ a simple 17-point Clenshaw-Curtis quadrature 
scheme~\cite{libSGJC} already suffices to evaluate the integrals to 
better than 11-digit accuracy.
Note
that, although the sinusoidal factor in the integrand of the original 
integral (\ref{OriginalIntegral}) exhibits 10$\times$ more rapid 
variation for $kR=1.0$ than for $kR=0.1$, the TDM reduction
to the 1D integrand
smooths this behavior to such an extent that the two cases are 
nearly indistinguishable in (\ref{CTExampleFigure}).

\begin{figure}
\begin{center}
\resizebox{0.5\textwidth}{!}{\includegraphics{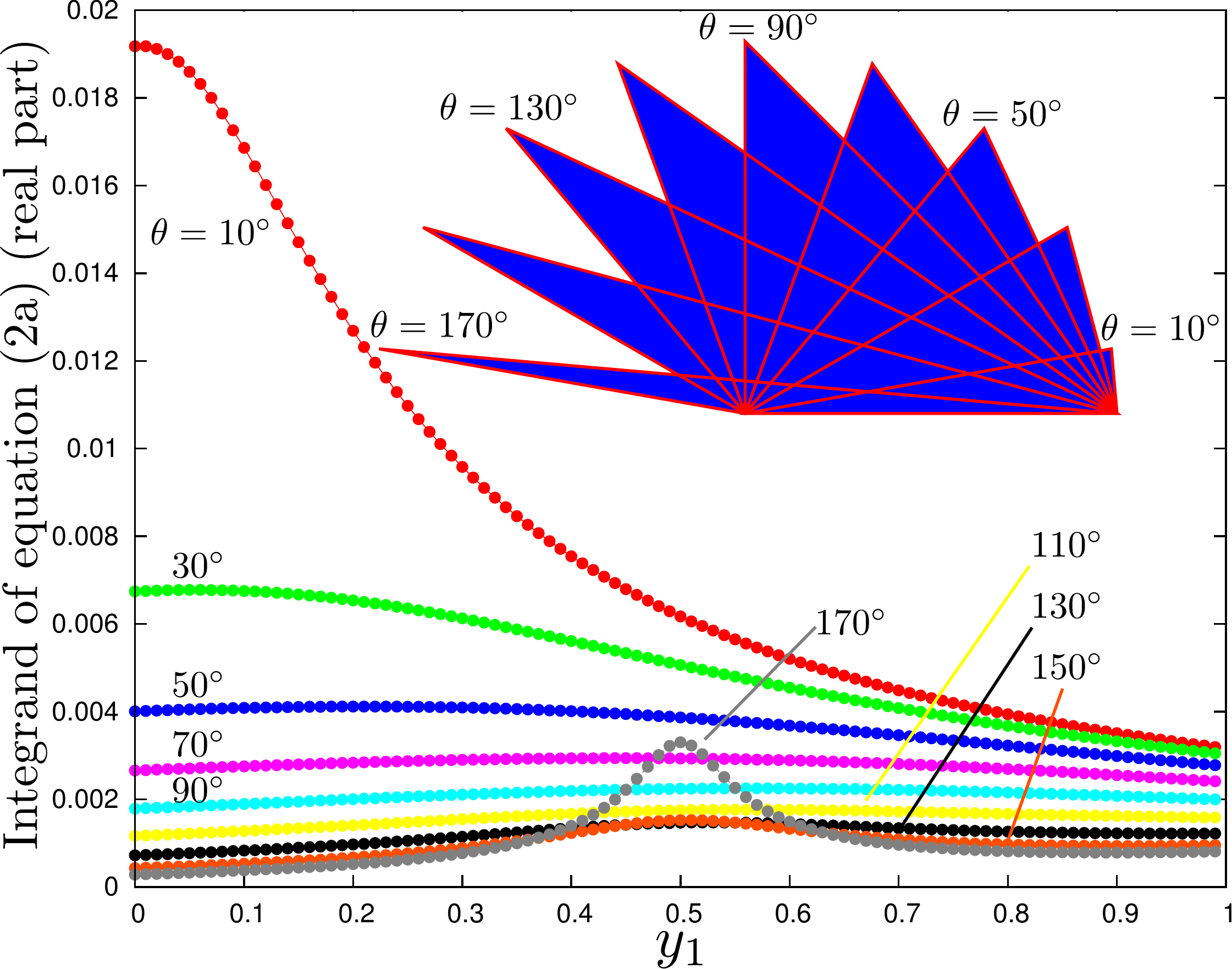}}
\caption{The real part of the integrand of equation (2a) for the case 
         $=P\supt{EFIE1}$ and kernel $K=K\supt{EFIE}$
         for a triangle in the $xy$ plane with vertices at  
         $(x,y)=(0,0), (L,0), (L\sin\theta,L\cos\theta)$
         for $L=0.1$ and 8 distinct values of $\theta$.
         In each case the wavenumber is $k=0.1/R$ with $R$ the
         triangle radius.
        }
\label{CTByThetaExampleFigure}
\end{center}
\end{figure}
How are these results modified for triangles of less-regular shapes?
Figure \ref{CTByThetaExampleFigure} plots the real part of the
integral of equation (2a), again for the choices
$\{P,K\}=\{P\sups{EFIE1},K\sups{EFIE}\},$ for a triangle in
the $xy$ plane with vertices
$(x,y)=(0,0), (L,0), (L\sin\theta,L\cos\theta)$ with $L=0.1$
and 8 distinct values of $\theta$. 
\begin{figure}[b]
\begin{center}
\resizebox{0.5\textwidth}{!}{\includegraphics{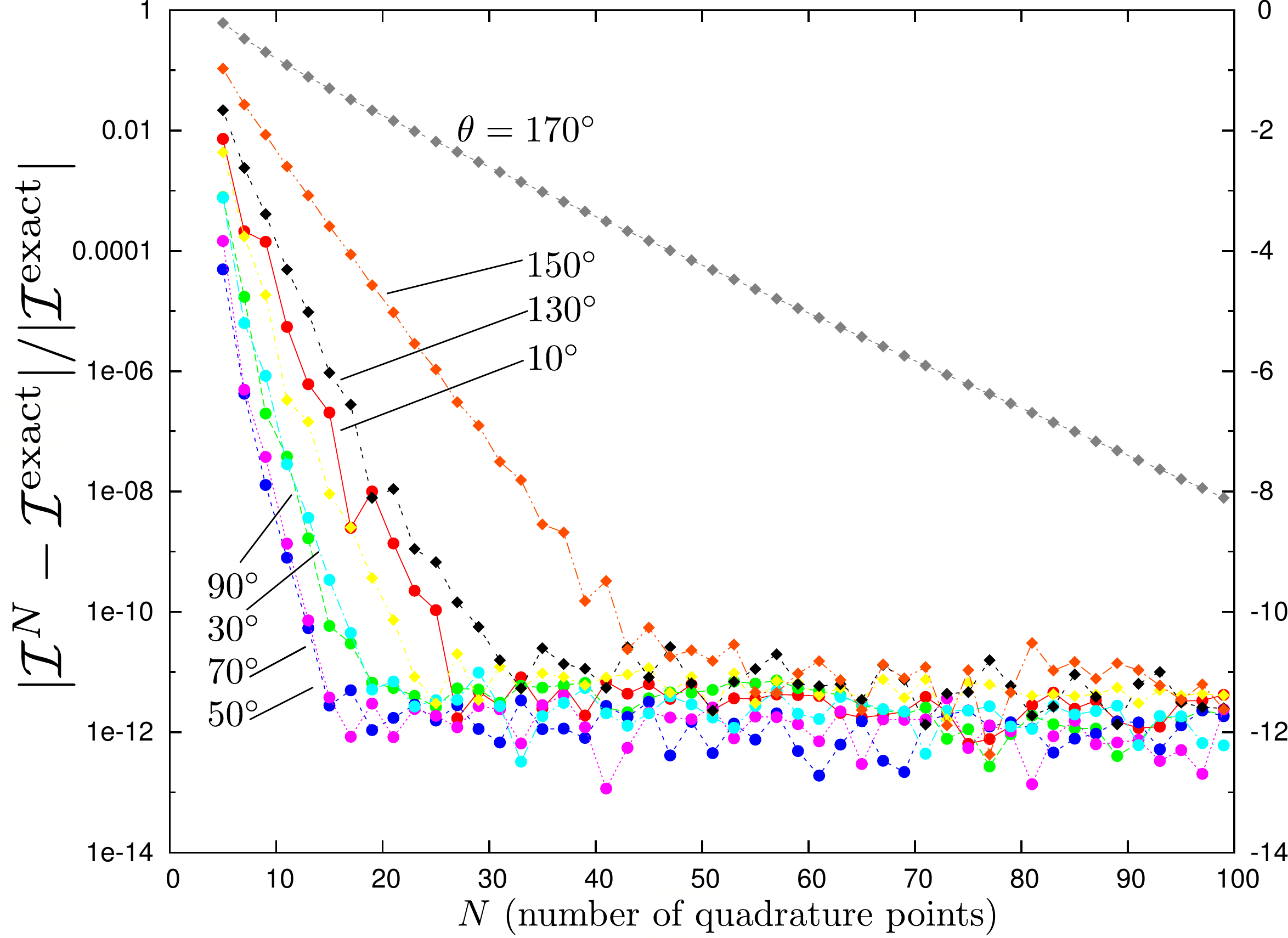}}
\caption{TDM convergence in the common-triangle case for a 
         once-integrable kernel. Plotted is the error vs. number 
         of Clenshaw-Curtis quadrature points incurred by
         numerical integration of the integrands
         plotted in Figure \ref{CTByThetaExampleFigure}.
         ($\mathcal{I}^N$ is the $N-$point Clenshaw-Curtis
          approximant to the integral.)
        }
\label{CTConvergenceFigure}
\end{center}
\end{figure}
The integrand exhibits slightly
more rapid variation in the extreme cases $\theta=10^\circ,170^\circ,$
but remains sufficiently smooth to succumb readily to low-order
quadrature. To quantify this, Figure \ref{CTConvergenceFigure}
plots, versus $N$, the relative error incurred by numerical
integration of the integrands of Figure \ref{CTByThetaExampleFigure}
using $N$-point Clenshaw-Curtis quadrature. 
(The relative error is defined as 
$|\mathcal{I}^N - \mathcal{I}\sups{exact}|/|\mathcal{I}\sups{exact}|$
where $\mc I^N$ and $\mc I\sups{exact}$ are the $N$-point 
Clenshaw-Curtis quadrature approximations to the integral
and the ``exact'' integral as evaluated by 
high-order \red{Clenshaw-Curtis} quadrature 
(with $N>100$ integrand samples per dimension).
For almost all cases 
we obtain approximately $12$-digit accuracy with just 20 to 30
quadrature points, with only the most extreme-aspect-ratio 
triangles exhibiting slightly slower convergence.
\red{
[Note that the $x$-axis quantity $N$ in Figure \ref{CTConvergenceFigure},
 as well as in 
 Figures \ref{CEConvergenceFigure1},
 \ref{CEConvergenceFigure2},
 and
 \ref{CVConvergenceFigure},
 represents
 the total number of integrand samples required to evaluate
 the \textit{full four-dimensional integral} (\ref{OriginalIntegral}), 
 not any lower-dimensional portion of this integral
 (such as might be the case for other integration schemes
 that---unlike the method of this paper---divide the original 
 integral into inner and outer ``source'' and ``target'' 
 integrals which are handled separately).
]
}

It is instructive to compare Figures \ref{CTByThetaExampleFigure} 
and \ref{CTConvergenceFigure} to Figure 1 
of~\citeasnoun{Taskinen2003}, which applied Duffy-transformation
techniques to the \textit{two-dimensional} integral of $1/r$
over a single triangle (in contrast to the four-dimensional
integrals over triangle pairs considered in this work).
With the triangle assuming various distorted shapes
similar to those in the inset of Figure (\ref{CTByThetaExampleFigure}), 
\citeasnoun{Taskinen2003} observed a dramatic slowing of the
convergence of numerical quadrature as the triangle aspect 
ratio worsened, presumably because the integrand exhibits 
increasingly rapid variations. In contrast, Figures
\ref{CTByThetaExampleFigure} and \ref{CTConvergenceFigure} 
indicate that no such catastrophic degradation in integrand 
smoothness occurs in the four-dimensional case, perhaps
because the analytical integrations effected by the 
TDM reduction from (\ref{OriginalIntegral}) to (2a)
smooth the bad integrand behavior that degrades convergence 
in the two-dimensional case. (Techniques for improving
the convergence of two-dimensional integrals over triangles
with extreme aspect ratios were discussed in~\citeasnoun{Khayat2008}.)

\subsection{Common-edge examples}

\begin{figure}
\begin{center}
\resizebox{0.5\textwidth}{!}{\includegraphics{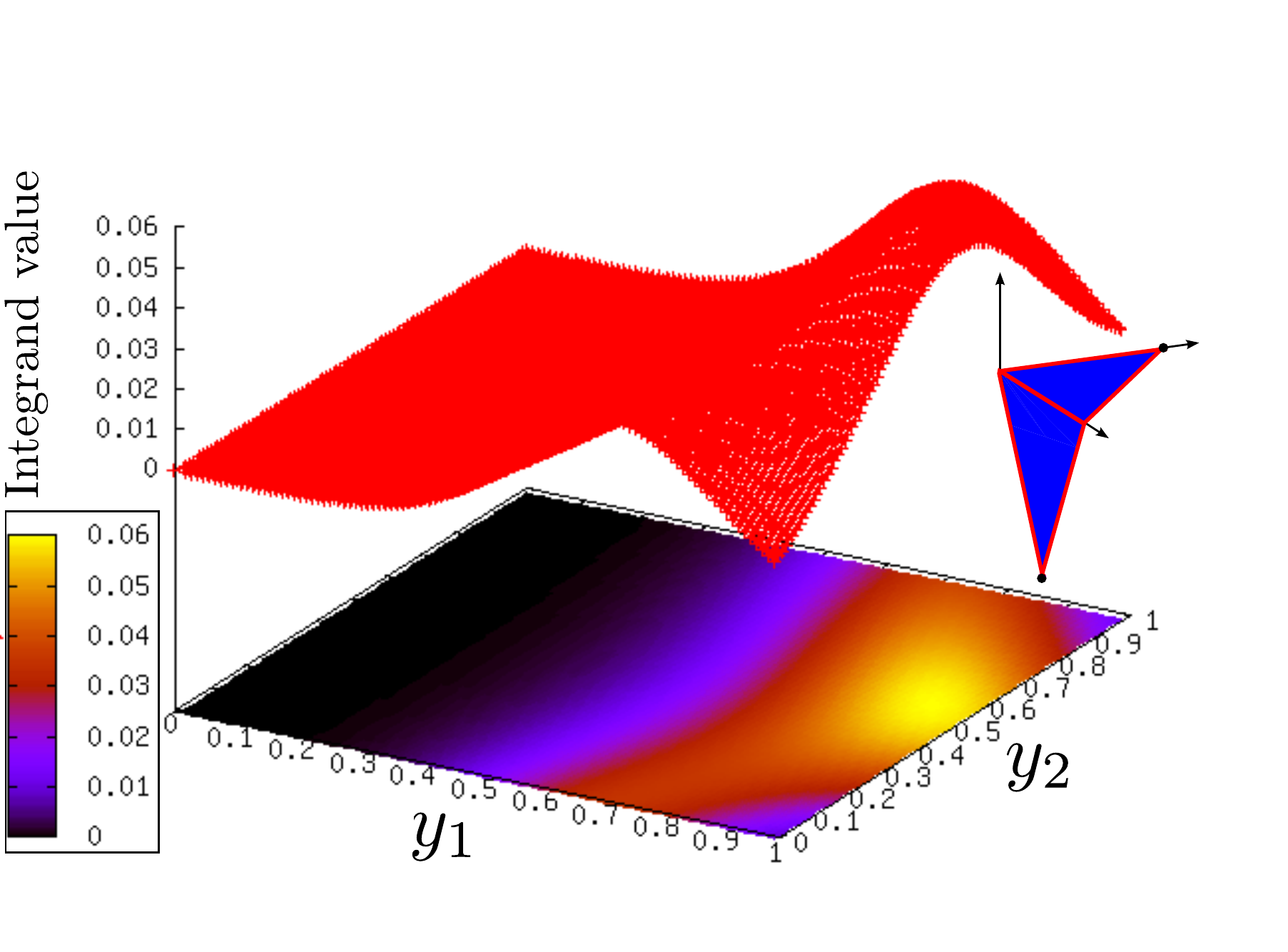}}
\caption{TDM integrand for a common-edge case with a 
once-integrable kernel. Plotted is the integrand of equation 
(2b) with polynomial $P=P\sups{MFIE}$ and kernel $K=K\sups{MFIE}$
for a right-angle common-edge panel pair (inset). }
\label{CEIntegrandFigure}
\end{center}
\end{figure}

\begin{figure}
\begin{center}
\resizebox{0.5\textwidth}{!}{\includegraphics{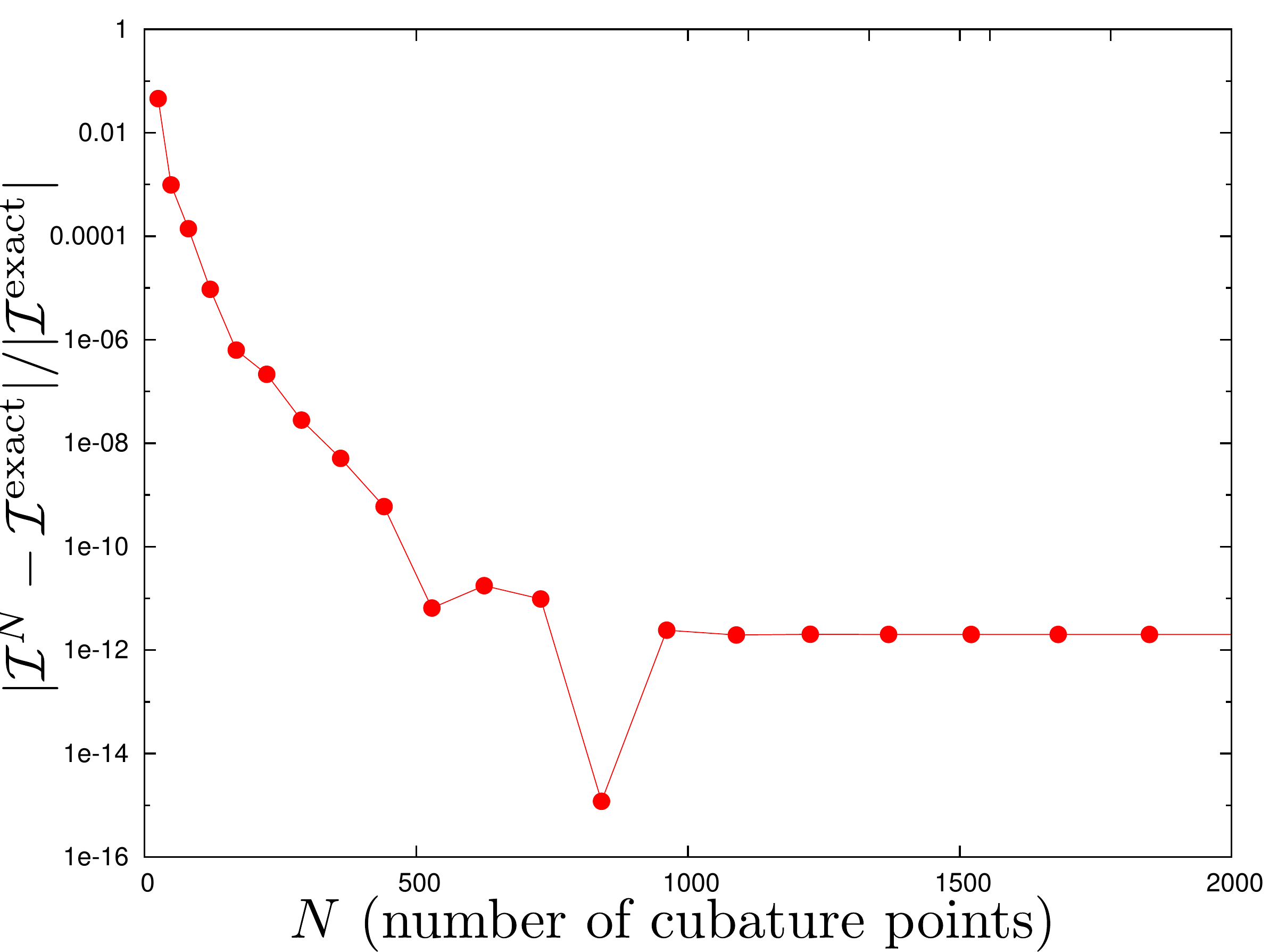}}
\caption{TDM convergence for a common-edge case with a 
         once-integrable kernel. Plotted is the error vs. 
         the total number of cubature points
         incurred by numerical integration of the integrand
         plotted in Figure \ref{CEIntegrandFigure}.
         The cubature scheme is simple two-dimensional
         nested Clenshaw-Curtis quadrature with equal numbers
         of quadrature points in each dimension.}
\label{CEConvergenceFigure1}
\end{center}
\end{figure}

\begin{figure}
\begin{center}
\resizebox{0.5\textwidth}{!}{\includegraphics{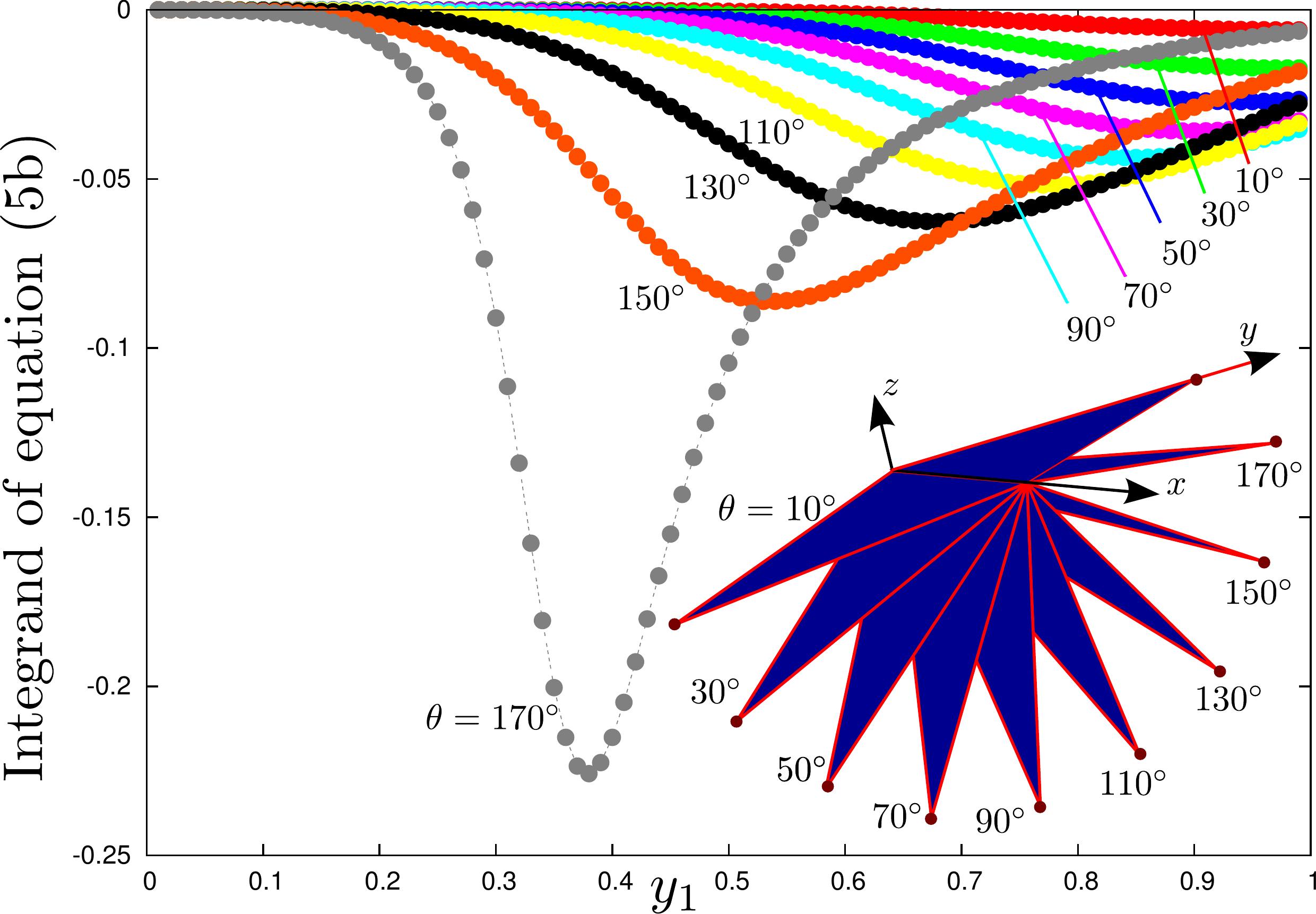}}
\caption{TDM integrand in a common-edge case with a twice-integrable
kernel. Plotted is the integrand of equation (5b) with polynomial
$P=P\supt{MFIE}$ and kernel $K(r)=\frac{1}{4\pi r^3}$ for 
common-edge panel pairs \red{ranging from nearly coplanar 
($\theta=10^\circ$) to right-angle ($\theta=90^\circ$)
to nearly coincident ($\theta=170^\circ$).}
In all cases the integrand is well-behaved and amenable
to low-order numerical quadrature (see Figure \ref{CEConvergenceFigure2}).
}
\label{CEVsThetaFigure}
\end{center}
\end{figure}

As an example of a common-edge case, Figure \ref{CEIntegrandFigure} plots 
the two-dimensional integrand of equation (2b) for the choice of polynomial
$P\sups{MFIE}=
 [ (\vb x-\vb Q) 
   \cdot 
   (\vb x^\prime-\vb Q^\prime)
 ] \cdot (\vb x-\vb x^\prime)$ 
and kernel $K\sups{MFIE}(r)=(ikr-1)\frac{e^{ikr}}{4\pi r^3}$, 
a combination which arises in the MFIE formulation 
with RWG functions (Section \ref{BEMFormulationsSection}).
The triangle pair (inset) is the right-angle pair 
$ \mathcal{T}=\{(0,0,0), (L,0,0), (0,L,0)\}$ 
and 
$ \mathcal{T}^\prime=\{(0,0,0), (L,0,0), (L/2,0,-L)\}$ 
with $L=0.1.$
The RWG source/sink vertices are indicated by black dots in the 
inset. 
The $k$ parameter in the Helmholtz kernel is chosen 
such that $kR=0.628$ where $R$ is the maximum panel radius.
The integrand is smooth and is amenable to straightforward
two-dimensional cubature. 
To quantify this, Figure \ref{CEConvergenceFigure1}
plots the error vs. number of cubature points incurred by 
numerical integration of the integrand plotted in Figure 
\ref{CEVsThetaFigure}. The cubature scheme is simply 
nested two-dimensional Clenshaw-Curtis cubature, with 
the same number of quadrature points per dimension.
Although the added dimension of integration inevitably
necessitates the use of more integration points than 
were needed in the 1D cases examined above, nonetheless
we achieve 12-digit accuracy with roughly 500 cubature 
points.

\begin{figure}
\begin{center}
\resizebox{0.5\textwidth}{!}{\includegraphics{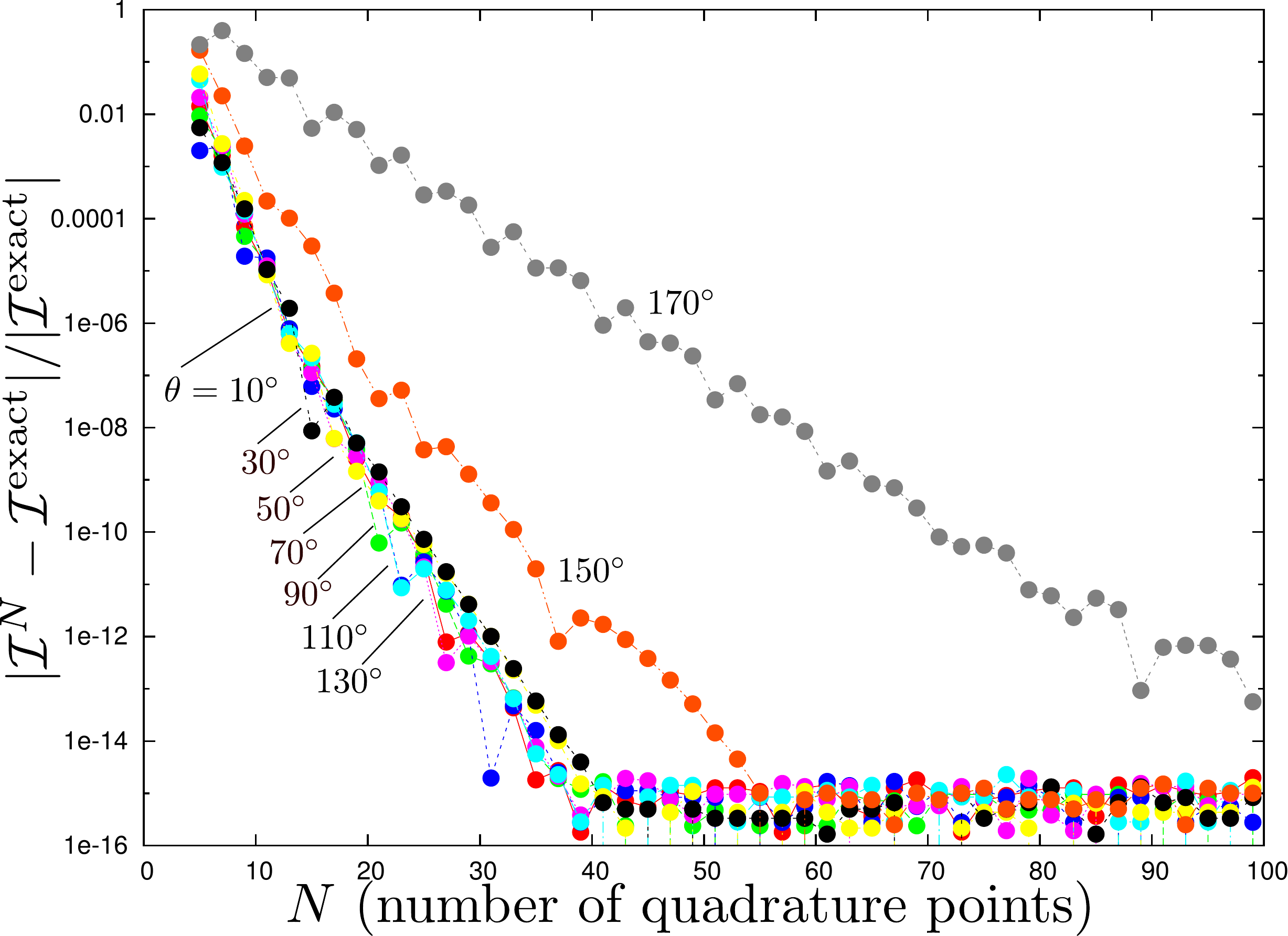}}
\caption{TDM convergence in a common-edge case for a 
         twice-integrable kernel. Plotted is the error vs. 
         number of quadrature points incurred by numerical integration 
         of the integrands plotted in Figure \ref{CEVsThetaFigure}.
         The cubature scheme is Clenshaw-Curtis quadrature.
         \red{
         The convergence rate is essentially independent of 
         the angle $\theta$ between the normals to the panels
         except for the near-singular case $\theta=170^\circ.$}
             }
\label{CEConvergenceFigure2}
\end{center}
\end{figure}

The kernel $K\sups{MFIE}(r)$ approaches $-1/(4\pi r^3)$ for small
$r$. (As noted above, pairing with $P\sups{MFIE}$ 
reduces the singularity of the overall integrand 
due to the vanishing of $P\sups{MFIE}$ at $r=0$.)
If we consider the integral of just this most singular
contribution---that is, if in (\ref{OriginalIntegral}) we retain 
the polynomial $P=P\sups{MFIE}$ but now replace 
the kernel $K\sups{MFIE}(r)$ with $K(r)=1/(4\pi r^3)$---then
we have a \textit{twice-integrable} kernel and the Taylor-Duffy 
reduction yields a \textit{one-dimensional} integral, equation 
(5b), instead of the two-dimensional integrand (2b) plotted 
in (\ref{CEIntegrandFigure}). 
Figure \ref{CEVsThetaFigure} plots the 1-dimensional integrand
obtained in this way for several common-edge panel pairs
obtained by taking the unshared vertex of $\mc T^\prime$ 
to be the point $(L/2,-L\cos\theta,-L\sin\theta)$
with $\theta$ ranging from $\theta=10^\circ$ to $170^\circ$.
(The inset of Figure \ref{CEIntegrandFigure} corresponds 
to $\theta=90^\circ.$)
For all values of $\theta$ the integrand is smooth
and readily amenable to numerical quadrature.
Figure \ref{CEConvergenceFigure2}
plots the convergence vs. number of quadrature points 
for numerical integration by Clenshaw-Curtis quadrature
of the integrands plotted in Figure \ref{CEVsThetaFigure}.
\red{
The convergence rate is essentially independent of $\theta$
throughout the entire range $\theta\in[0,130^\circ].$
}

\subsection{Common-vertex example}
\begin{figure}
\begin{center}
\resizebox{0.5\textwidth}{!}{\includegraphics{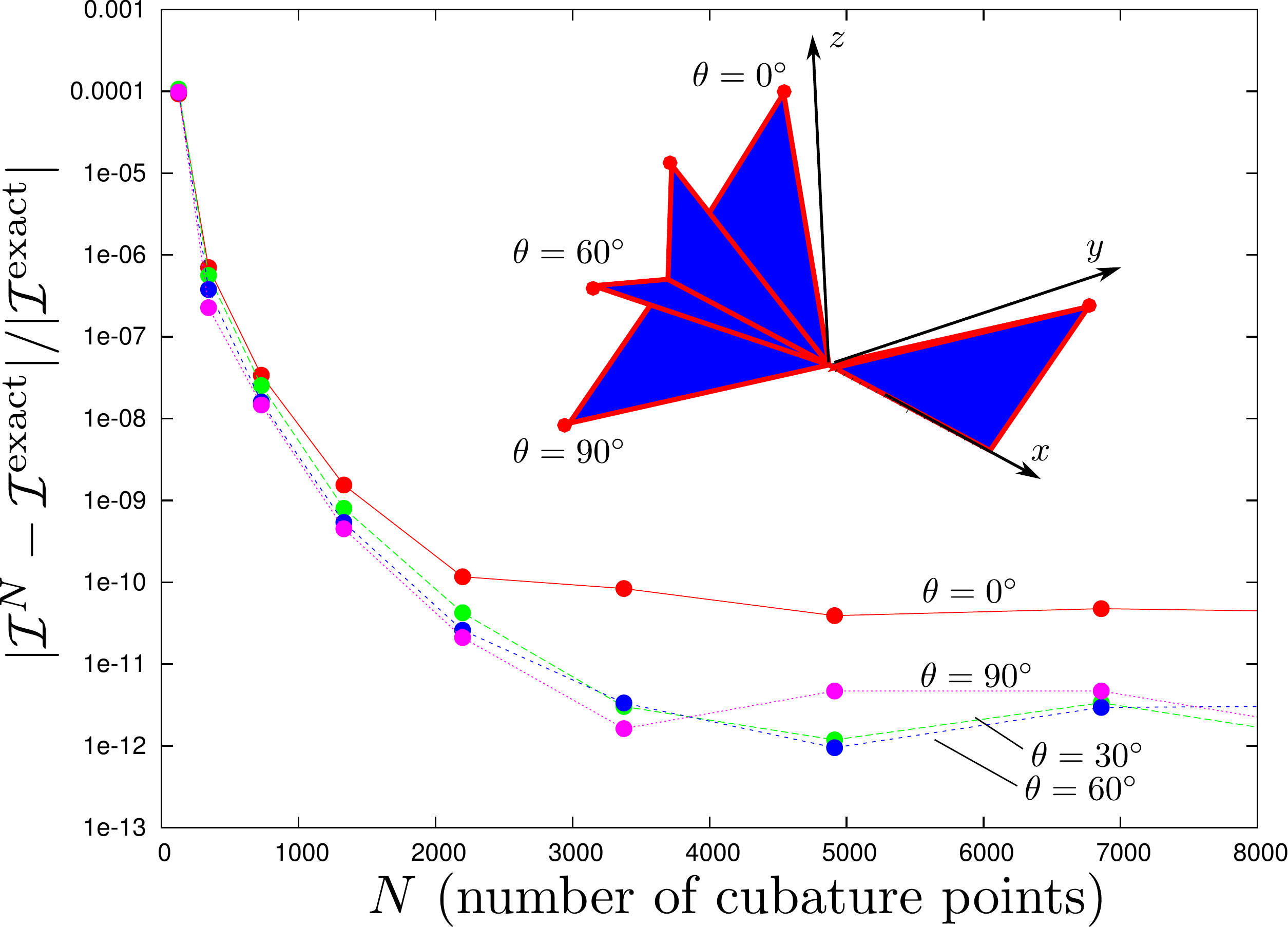}}
\caption{TDM convergence in a common-vertex case for a 
         once-integrable kernel.
         Plotted is the error vs. number of cubature points
         incurred by numerical integration of Equation 2(c)
         for the common-vertex triangle pairs shown in the inset
         (see text) and polynomial/kernel pair 
         $\{P,K\}$=$\{P\supt{EFIE1}, K\supt{EFIE}\}$.
         The convergence rate is essentially independent of 
         the angle $\theta$.}
\label{CVConvergenceFigure}
\end{center}
\end{figure}

As an example of a common-vertex case, Figure \ref{CVConvergenceFigure} plots 
the convergence of equation (2c) for 
$\{P,K\}$=$\{P\sups{EFIE1}, K\sups{EFIE}\}$,
the same pair considered for the common-triangle example above.
The triangle pair (inset) is 
$ \mathcal{T}=\{(0,0,0), (L,0,0), (L^\prime,L,0)\}$ 
and 
$ \mathcal{T}^\prime=\{(0,0,0), (-L,0,0), (-L^\prime\sin\theta,L^\prime\cos\theta)\}$
with $\{L,L^\prime\}=\{0.1,0.02\}$ and various values of $\theta$.
The $k$ parameter in the Helmholtz kernel is chosen 
such that $kR=0.628$ where $R$ is the maximum panel radius.
RWG source/sink vertices are indicated by dots in the 
inset. 
The figure
plots the error vs. number of cubature points incurred by 
numerical integration of the integrand plotted in Figure 
\ref{CEVsThetaFigure}. The cubature scheme is simply 
nested three-dimensional Clenshaw-Curtis cubature, with 
the same number of quadrature points per dimension.
This is probably not the most efficient cubature scheme
for a three-dimensional integral, but the figure demonstrates 
that the error decreases steadily and rapidly with the 
number of cubature points.
The convergence rate is essentially independent of $\theta$.

\section{Application to Full-Wave BEM Solvers: 
         Evaluation and Caching of Series-Expansion Terms}
\label{CachingSection}

We now take up the question of how the generalized
TDM may be most effectively deployed in practical
implementations of full-wave BEM solvers using RWG
basis functions.
To assemble the BEM matrix for, say, the 
PMCHWT formulation at a single frequency for a
geometry discretized into $N$ triangular surface panels,
we must in general compute $\approx\, 12N$ singular
integrals of the form (\ref{OriginalIntegral}) with
roughly $\{2N,4N,6N\}$ instances of the
common-\{triangle, edge, vertex\} cases. [For each
panel pair we must compute integrals involving two
separate kernels ($K\sups{EFIE}$ and
$K\sups{MFIE}$), and neither of these kernels is
twice-integrable except in the short-wavelength limit.]

A first possibility is simply to evaluate all singular
integrals using the basic TDM scheme outlined in 
Section \ref{GeneralizedTaylorDuffySection}---that is,
for an integral of the form (\ref{OriginalIntegral})
with polynomial $P(\vb x, \vb x^\prime)$ and 
kernel $K(r)$ we write simply
\numeq{FullEvaluation}
{
  \mathcal{I}
   =
   \underbrace{\iint P(\vb x, \vb x^\prime) K(r) d\vb x^\prime d\vb x.}
            _{\text{evaluate by TDM}}
}
This method already suffices to evaluate all singular integrals
and would consitute an adequate solution for a medium-performance 
solver appropriate for small-to-midsized problems. However,
although the unadorned TDM successfully neutralizes 
singularities to yield integrals amenable to simple 
numerical cubature, we must still \textit{evaluate} 
those 1D-, 2D-, and 3D cubatures, and this task, even 
given the non-singular integrands furnished by the TDM, 
remains too time-consuming for the online stage of a 
high-performance BEM code for large-scale probems.

\red{
Instead, we propose here to accelerate the bare
TDM using singularity-subtraction (SS)
techniques~\cite{Taskinen2003, Jarvenpaa2006, TongChew2007}.
Subtracting from $K(r)$ the first $M$ terms in its
small-$r$ series expansion, we write
\numeq{KDecomposition}
{K(r)=\sum_{m=0}^{M-1} C_m r^{m-L} + K\sups{NS}(r)}
where $K\sups{NS}$ is nonsingular. Here $L$
is the most singular power of $1/r$
in the small-$r$ expansion of $K(r)$; for 
$\{K\sups{EFIE}, K\sups{MFIE}\}$ we have $L=\{1,3\}$.
(As noted below, we typically choose $M>L$, i.e. we 
subtract more than the minimal number of terms required
to desingularize the kernel.)
Equation (\ref{FullEvaluation}) becomes
\begin{align}
\mathcal{I}
&= \sum_{m=0}^{M-1} C_m
  \underbrace{\iint P(\vb x, \vb x^\prime) r^{m-L} d\vb x^\prime d\vb x}
            _{\text{evaluate by TDM}}
\label{MSubEval} \\
&\qquad +\underbrace{\iint P(\vb x, \vb x^\prime) 
                    K\sups{NS}(r)d\vb x^\prime d\vb x}
            _{\text{evaluate by cubature}}
\nonumber
\end{align}
The first set of terms on the RHS of (\ref{MSubEval}) involve 
integrals of the form (\ref{OriginalIntegral}) with kernels $r^p$,
which we evaluate using the generalized TDM proposed
in this paper. The last term is a nonsingular integral which we 
evaluate using simple low-order numerical cubature.
}
Such a hybrid TDM-SS approach has several advantages.
\textbf{(a)} The integrals involving the $r^p$ kernel
are \textit{independent of frequency and material properties},
even if $K(r)$ depends on these quantities through the wavenumber
$k=\sqrt{\epsilon\mu}\cdot\omega$. [The $k$ dependence of the first
set of terms in (\ref{MSubEval}) is contained entirely in the 
constants $\{C_m\}$, which enter only as multiplicative 
prefactors outside the integral sign.]
This means that we need only compute these integrals \textit{once} for a 
given geometry, after which they may be stored and reused 
many times for computations at other frequencies or for
scattering geometries involving the same shapes but 
different material properties. (The caching and reuse of 
frequency-independent contributions to BEM integrals has 
been proposed before~\cite{Taskinen2003}.)
\textbf{(b)} The $r^p$ kernels in the first set of integrals 
are \textit{twice-integrable}. This means that the improved 
Taylor-Duffy scheme discussed in Section 
\ref{TwiceIntegrableKernelSection} is available, 
significantly accelerating the computation;
for the common-triangle case these integrals maybe
evaluated in closed analytical form.
\textbf{(c)} The $M$ integrals in the first set
all involve the \textit{same} $P$ polynomial. This
means that the computational overhead required
to evaluate TDM integrals involving this polynomial need
only be done once and may then be reused for all $M$ 
integrals. Indeed, all of the integrals on the first line
of (\ref{MSubEval}) may be evaluated simultaneously
as the integral of an $M$-dimensional vector-valued
integrand; in practice this means that the cost of
evaluating all $M$ integrals is nearly independent
of $M$.
\textbf{(d)} Because $K\sups{NS}$ has been relieved
of its most rapidly varying contributions,
it may be integrated with good accuracy by a simple
low-order cubature scheme. \{We evaluate the 4-dimensional 
integral in the last term of (\ref{MSubEval})
using a 36-point cubature rule~\cite{Cools2003}.\}

%
\begin{figure}
\begin{tabular}{c}

\resizebox{0.48\textwidth}{!}{\includegraphics{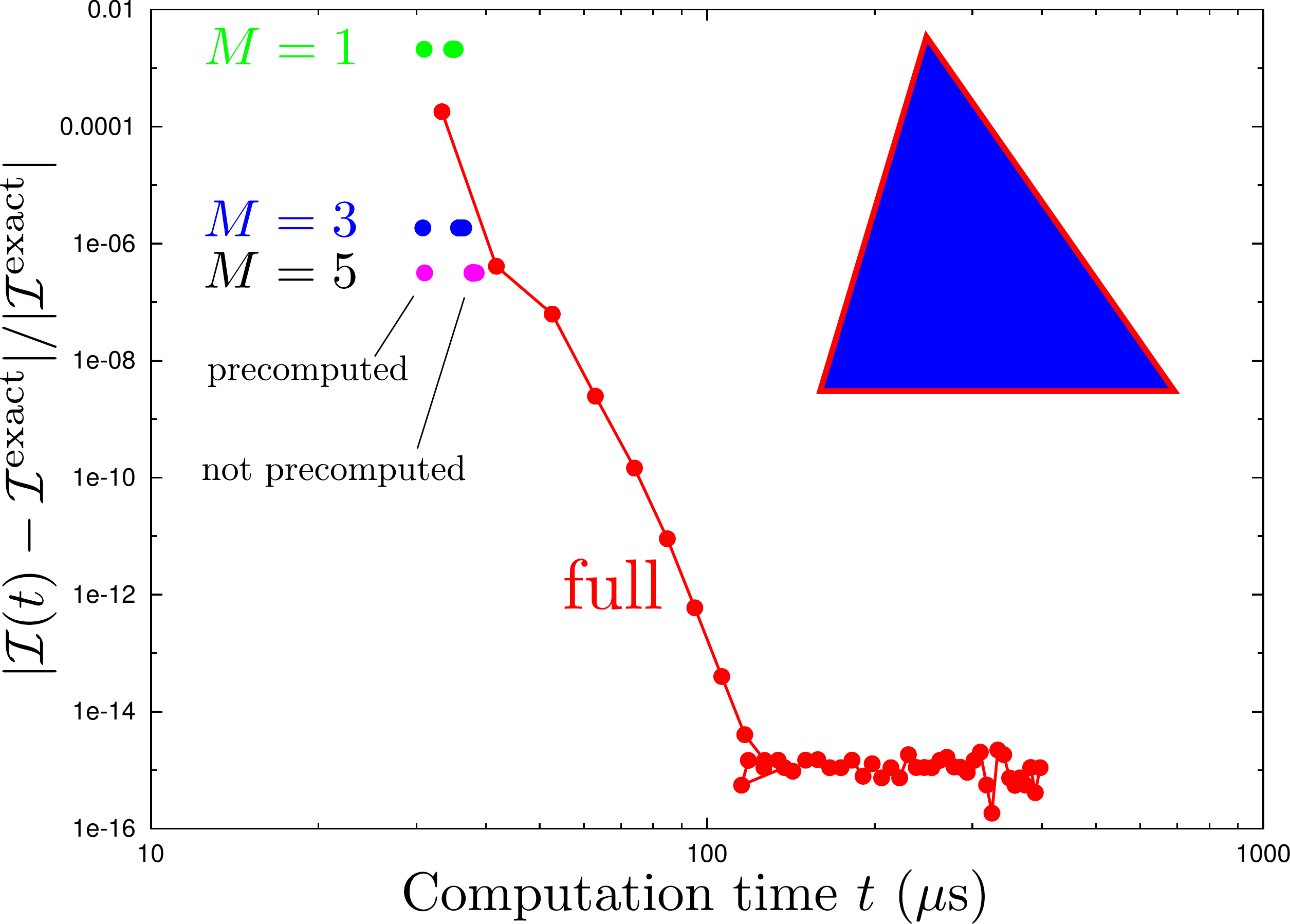}}
\\
\resizebox{0.48\textwidth}{!}{\includegraphics{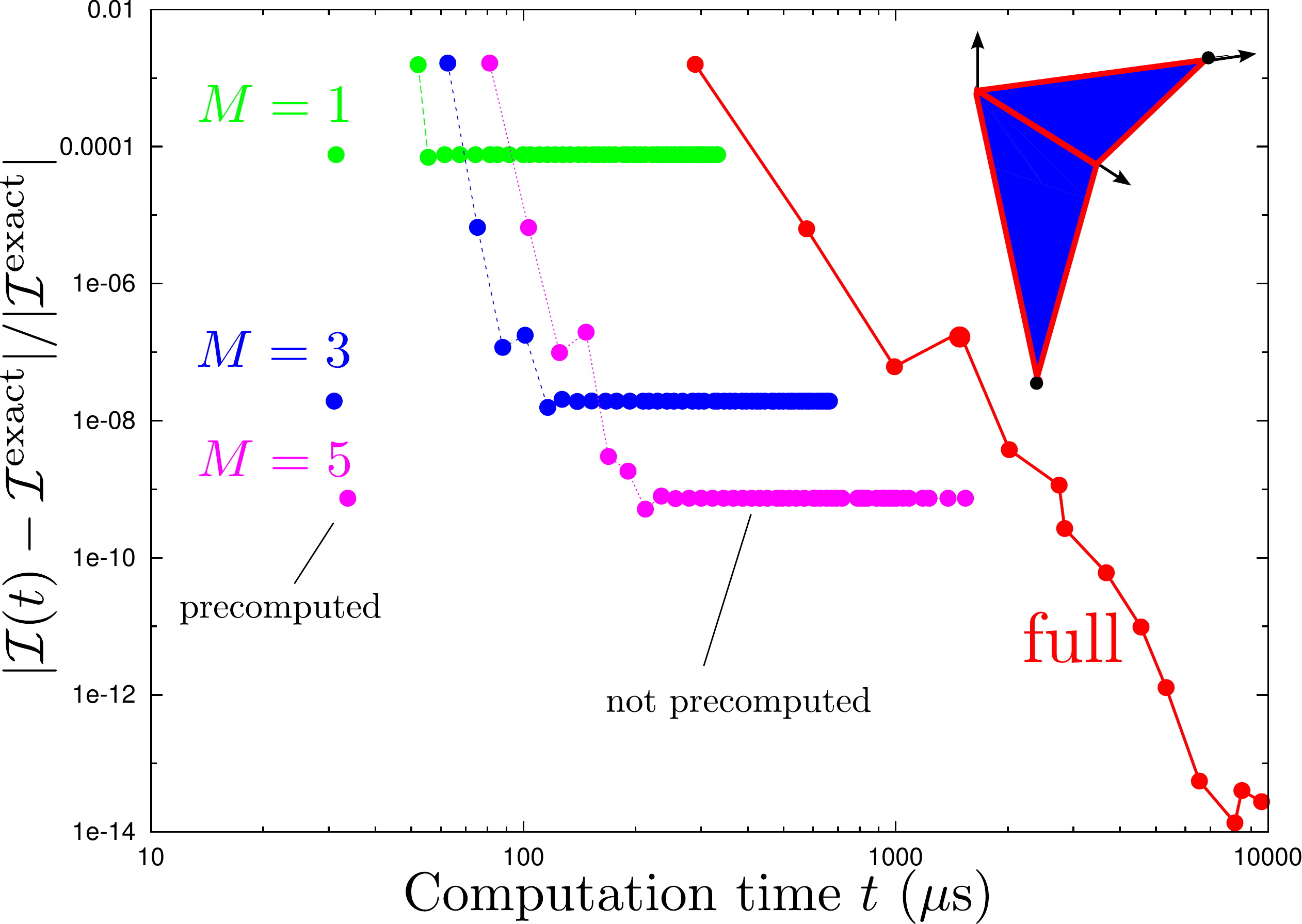}}
\\
\resizebox{0.48\textwidth}{!}{\includegraphics{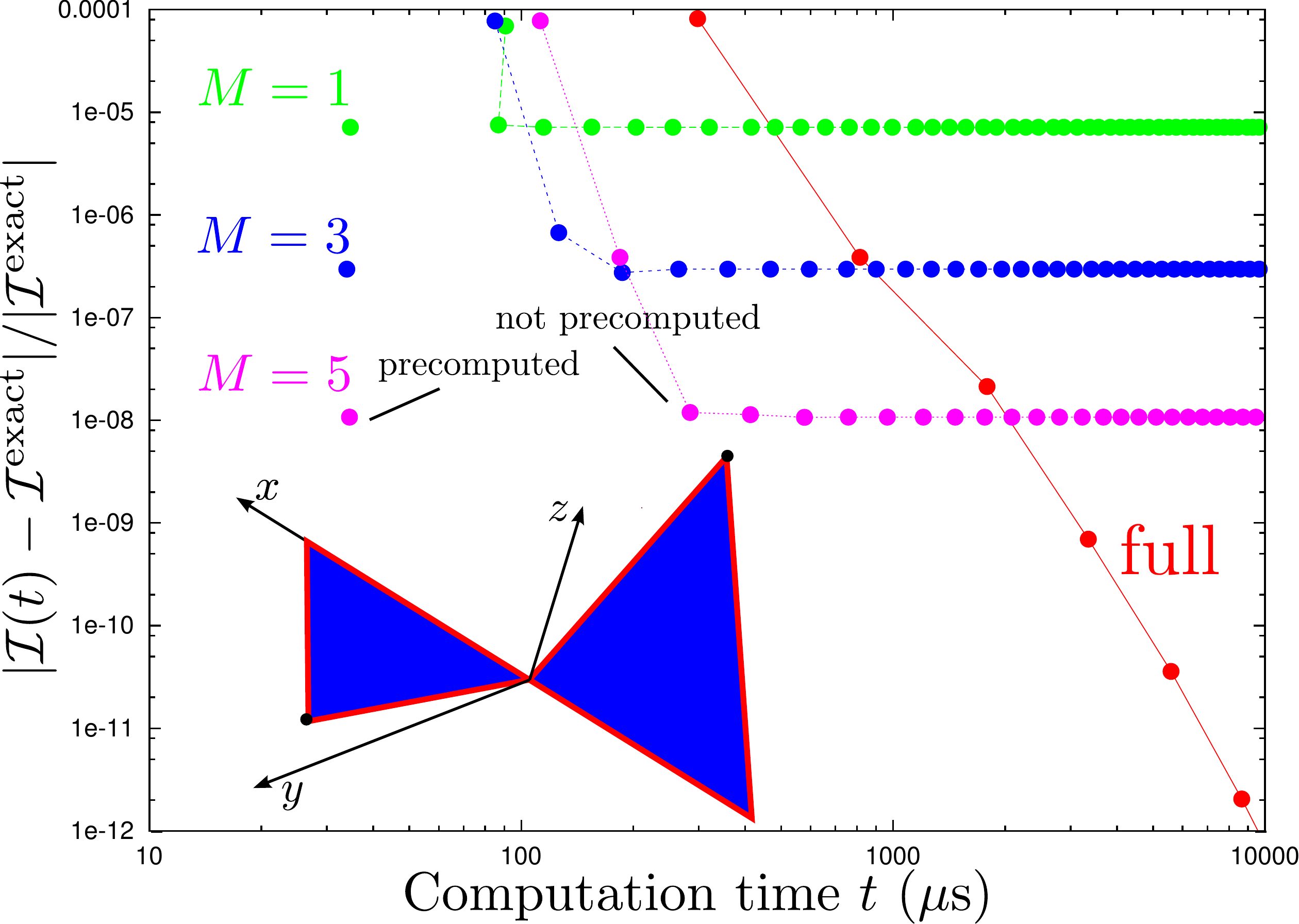}}
\end{tabular}
\caption{Accuracy vs. computation time for approaches to
         evaluating singular integrals.
         \textit{Top:} Common-triangle case with the 
         triangle of Figure \ref{CTExampleFigure}.
         \textit{Center:} Common-edge case with the 
         triangle of Figure \ref{CEIntegrandFigure}.
         \textit{Bottom:} Common-vertex case with the 
         $\theta=30^\circ$ triangle of Figure \ref{CVConvergenceFigure}.
         In each case the integral considered is 
         (\ref{OriginalIntegral})
         with $\{P,K\}=\{P\supt{EFIE1},K\supt{EFIE}\}$
         (Section \ref{BEMFormulationsSection}).
         The wavenumber $k$
         is chosen such that $kR=0.628$ where $R$ is the maximum
         panel radius, i.e. the linear size of the panel
         is approximately 1/10 the wavelength.
         Curves marked ``full'' correspond to equation
         (\ref{FullEvaluation}), i.e. numerical cubature
         of the 1, 2, or 3-dimensional integral in equation 
         (2). Curves marked $M=\{1,3,5\}$ correspond to 
         the singularity-subtraction scheme of equation
         (\ref{MSubEval}); the singular integrals
         are evaluated by numerical cubature of the 
         0, 1, or 2-dimensional integral in equation (5),
         while the remaining (non-singular) integral is 
         evaluated using a 36-point four-dimensional 
         numerical cubature scheme.
         All TDM integrals are evaluated using 
         Clenshaw-Curtis quadrature (nested for 
         2- and 3-dimensional integrals).
        }
\label{ErrorVsTimeFigure}
\end{figure}

Figure \ref{ErrorVsTimeFigure} compares accuracy vs.
computation time for the methods of 
equation (\ref{FullEvaluation}) and equation 
(\ref{MSubEval}) with $M=\{1,3,5\}$
subtracted terms. The \{top, center, bottom\}
plots are for the \{CT, CE, CV\} cases using
the triangle pairs of Figures 
\{\ref{CTExampleFigure}, \ref{CEIntegrandFigure}, 
  \ref{CVConvergenceFigure}\}
(we choose the $\theta=30^\circ$ triangle pair
for the CV case). The integral computed 
is (\ref{OriginalIntegral}) with 
$\{P,K\}=\{P\sups{EFIE1},K\sups{EFIE}\}$.
The wavenumber $k$ is chosen such that $kR=0.628$ 
where $R$ is the maximum panel radius, so that
the linear size of the panels is approximately 
1/10 the wavelength.

The curves marked ``full'' in each plot 
correspond to equation (\ref{FullEvaluation}),
i.e. full evaluation by numerical cubature of 
the (1,2,3)-dimensional integral of equation (2).
The $M=\{1,3,5\}$ data correspond to 
equation (\ref{MSubEval}). 
For each $M$ value, the data point furthest to the
left is for the case in which we 
precompute the contributions of the singular 
integrals, so that the only computation time
is the evaluation of the fixed-order cubature.
The other data points for each $M$ value 
include the time incurred for numerical
cubature of equation (5) for the subtracted
(singular) terms at varying degrees of accuracy.
Beyond a certain threshold computation time
the integrals have converged to accurate
values, whereupon further computation time 
does not improve the accuracy with which we
compute the overall integral (because we 
use a fixed-order cubature for the
nonsingular contribution). Of course,
one could increase the cubature order for
the fixed-order contribution at the
expense of shifting all $M=\{1,3,5\}$
data points to the right.

In the common-triangle case, the integrals of the 
singular terms may be done in closed form 
[equation (5a)], so the data points marked 
``not precomputed'' all correspond to roughly 
the same computation time.
In the other cases, the ``not precomputed'' data points
for various values of the computation time correspond
to evaluation of the integrals (5b) or (5c) via
numerical cubature with varying numbers of cubature 
points.

Absolute timing statistics are of course 
heavily hardware- and implementation-dependent (in this case 
they were obtained on a standard desktop workstation),
but the picture of \textit{relative} timing that emerges
from Figure \ref{ErrorVsTimeFigure} is essentially 
hardware-independent. 
For a given accuracy, the singularity-subtracted
scheme (\ref{MSubEval}) is typically an order of
magnitude faster than the full scheme (\ref{FullEvaluation}), 
and this is true even if we include the time
required to compute the integrals of the 
frequency-independent subtracted terms.
If we \textit{precompute} those integrals,
the singularity-subtraction scheme is 
several orders of magnitude faster than the 
full scheme. For example, to achieve 
8-digit accuracy in the common-vertex case
takes over 2 ms for the full scheme 
but just 30 $\mu$s for the precomputed $M=5$
singularity-subtraction scheme.

The speedup effected by singularity subtraction is 
less pronounced in the common-triangle case. This
is because the full TDM integral (\ref{FullEvaluation})
is only 1-dimensional in that case and thus already
quite efficient to evaluate.

\section{Helmholtz-Kernel Integrals in the Short-Wavelength Limit}

\label{HighKSection}

The kernels $K\sups{EFIE}(r)=\frac{e^{ikr}}{4\pi r}$
and 
$K\sups{MFIE}(r)=(ikr-1)\frac{e^{ikr}}{4\pi r^3}$
become twice-integrable in the limit
$\text{Im }k\to \infty.$
More specifically, as shown in Appendix
\ref{FirstSecondIntegralAppendix}, the first integral
[equation (\ref{FirstIntegral})]
of these kernels takes the form 
$Q_1(r) + e^{ikX(r)}Q_2(r)$, where $Q_1,Q_2$
are Laurent polynomials in $r$
and $X$ is a nonvanishing quantity bounded below 
by the linear size of the triangles. 
When $\text{Im }k$ is large, 
the exponential factor makes the second term negligible,
and we are left with just $Q_1(r)$---which, as a 
sum of integer powers of $r$---is twice-integrable.
This means that the TDM-reduced version of 
integral (\ref{OriginalIntegral}) involves one 
fewer dimension of integration than in the 
usual case, i.e. we have equations (5) instead 
of equations (2). In particular, for the common-triangle
case the full integral may be evaluated in closed
form [equation (5a)].
\begin{figure}
\begin{center}
\begin{tabular}{c}
\resizebox{0.48\textwidth}{!}{\includegraphics{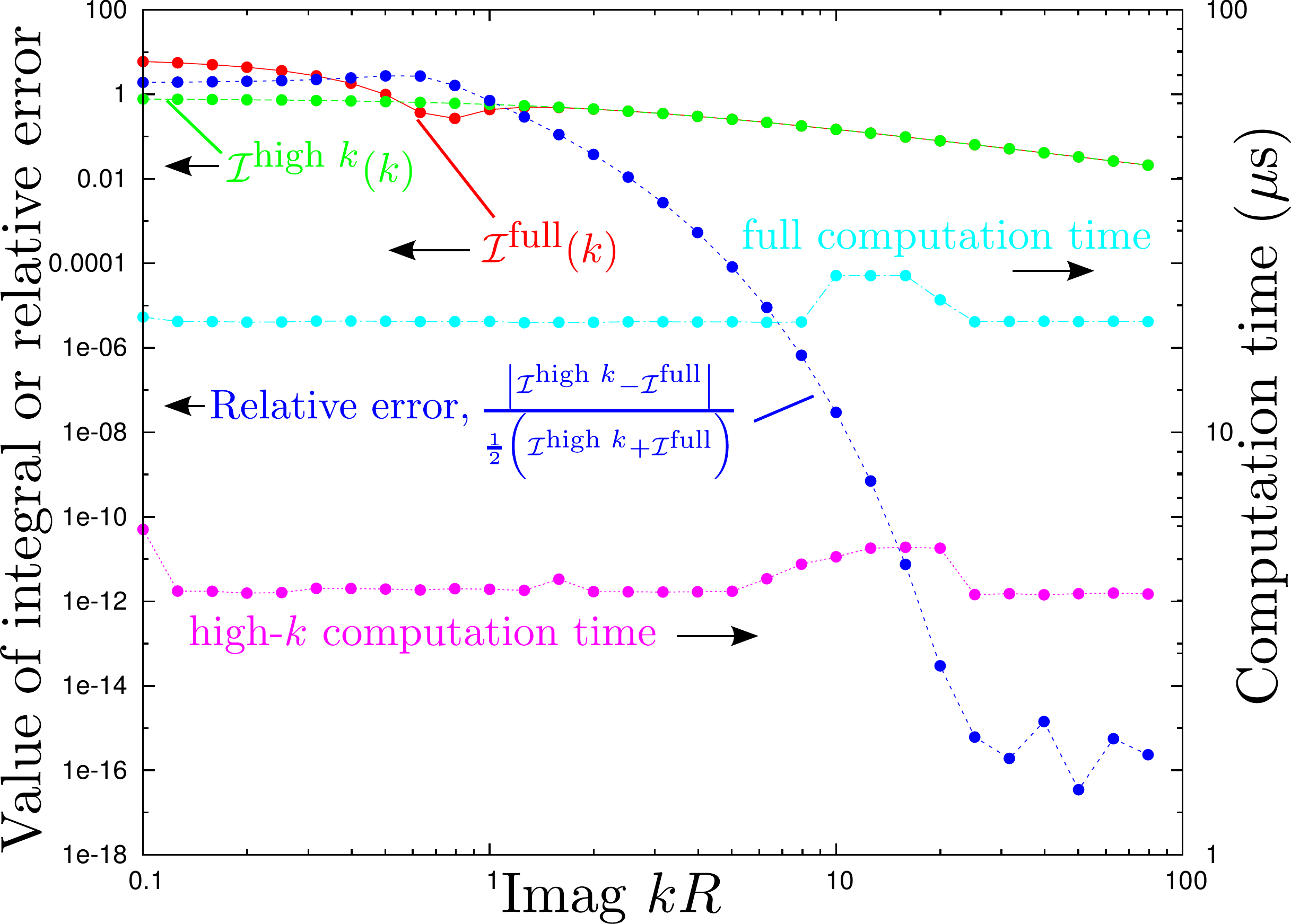}}
\\
\resizebox{0.48\textwidth}{!}{\includegraphics{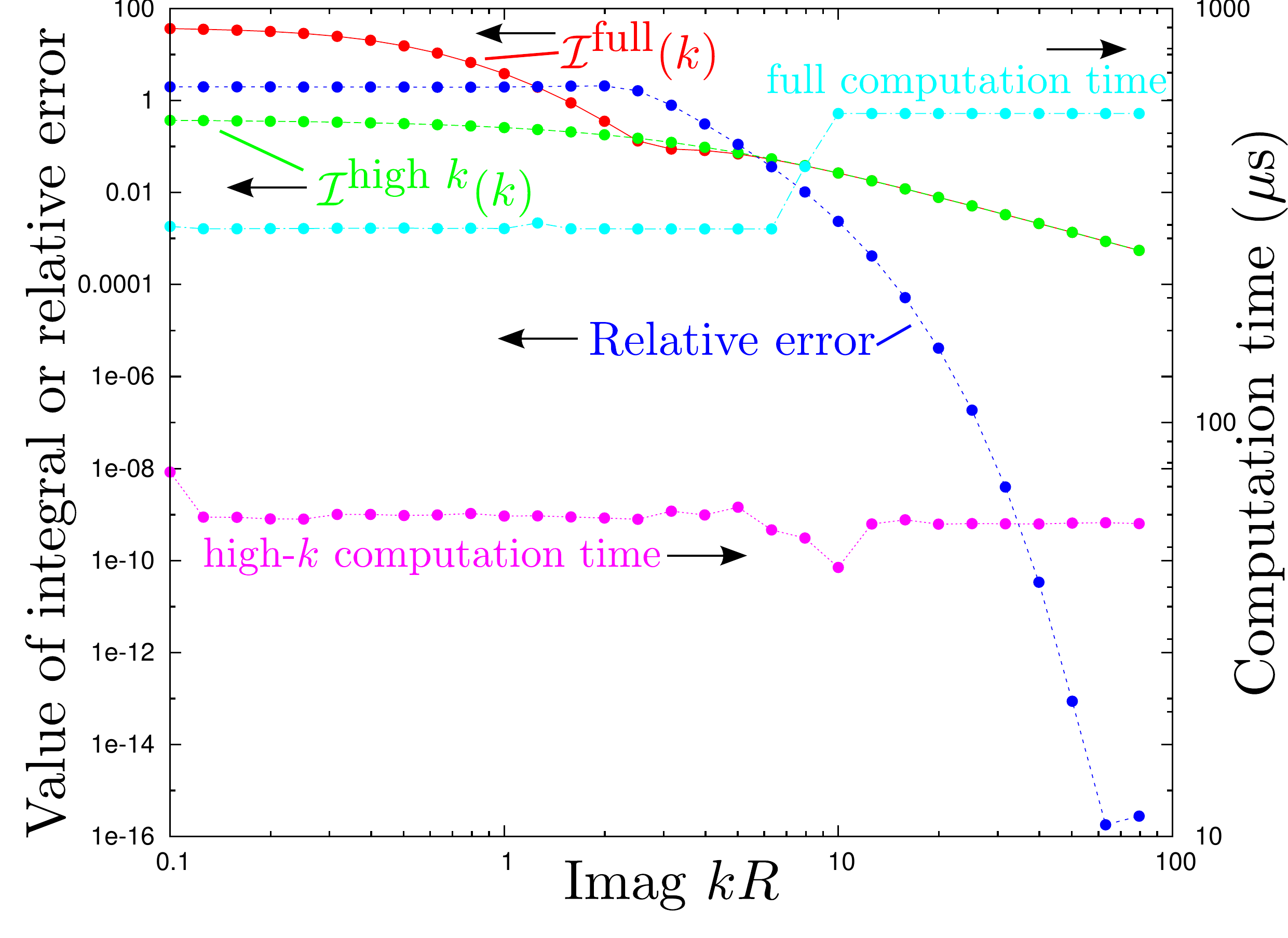}}
\end{tabular}
\end{center}
\caption{Accuracy and computation time for evaluation of integral
         (\ref{OriginalIntegral}) for 
         $\{P,K\}=\{1,\frac{e^{ikr}}{4\pi r}\}$
         via equations (2) and (5).
         \textit{Upper plot:} Common-triangle case for the triangle
         of Figure \ref{CTExampleFigure}.
         \textit{Lower plot:} Common-edge case for the triangle
         pair of Figure \ref{CEIntegrandFigure}.
         The $x$-axis measures the imaginary part of the 
         wavenumber $k$ in the Helmholtz kernel.
         The real part of $k$ is fixed at Re $kR=0.628$
         where $R$ is the maximum panel radius.
         In each case, the red curve is the value of the         
         $\{1,2\}$-dimensional integral computed using 
         the ``full'' equation (2),
         while the green curve is the value of the         
         $\{0,1\}$-dimensional integral computed using
         the ``high-$k$'' approximation to the integral
         involving equation (5).
         The blue curve is the relative error
         between the two calculations (their difference
         divided by their average).
         The cyan and magenta curves, respectively
         plot the (wall-clock) time required to 
         evaluate the integrals via the ``full''
         and ``high-$k$'' schemes. 
        }   
\label{HighKFigure}
\end{figure}

Figure \ref{HighKFigure} 
plots, for the common-triangle case of 
Figure \ref{CTExampleFigure} (upper plot) 
and the common-edge case of Figure \ref{CEIntegrandFigure}
(lower plot),
values and computation times for the integral
$\iint \frac{e^{ikr}}{4\pi r}\,d^4 r$---that is, 
equation (\ref{OriginalIntegral}) for the choices
$\{P,K\}=\{1,K\sups{EFIE}\}$---as evaluated
using the ``full'' scheme of equation (2) (red curves)
and using the ``high-$k$ approximation'' of 
equation (5) (green curves), with the exponentially-decaying
term in the first integral of the kernel neglected
to yield a twice-integrable kernel. The blue curves
are the relative errors between the two calculations.
The cyan and magenta curves respectively plot 
the (wall-clock) time required to compute the
integrals using the full and high-$k$ methods.
The high-$k$ calculation is approximately
one order of magnitude more rapid than the 
full calculation and yields results in good agreement
with the full calculation for values of Imag $kR$
greater than 10 or so.

\section{Conclusions}
\label{ConclusionsSection}
\begin{figure*}
\begin{center}
\resizebox{\textwidth}{!}{\includegraphics{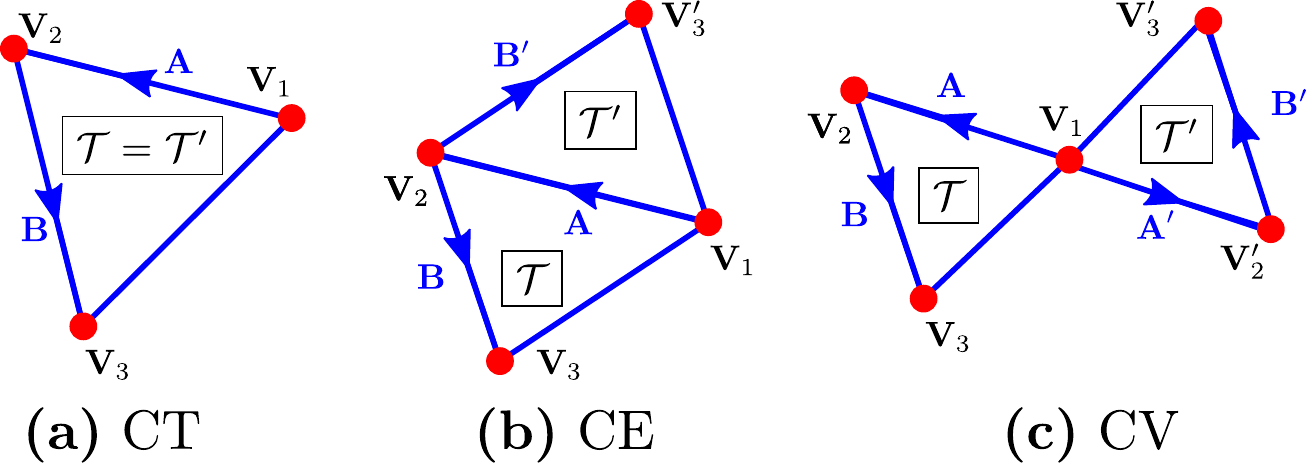}}
\caption{Labeling of triangle vertices and edges. 
         \textbf{(a)} The common-triangle case.
         \textbf{(b)} The common-edge case.
         \textbf{(c)} The common-vertex case.}
\label{VertexEdgeLabelFigure}
\end{center}
\end{figure*}

The generalized Taylor-Duffy method we have presented
allows efficient evaluation of a broad family of 
singular and non-singular integrals over triangle-pair
domains. The generality of the method allows a single
implementation ($\sim$1,500 lines of C++ code) to 
handle \textit{all} singular integrals arising in
several different BEM formulations, including 
electrostatics with triangle-pulse basis functions 
and full-wave electromagnetism with RWG basis 
functions in the EFIE, MFIE, PMCHWT, and N-M\"uller 
formulations.
In particular, for N-M\"uller integrals the method
presented here offers an alternative to the line-integral 
scheme discussed in \citeasnoun{Taskinen2003}. 
In addition 
to deriving the method and discussing 
practical implementation details, we presented several
computational examples to illustrate its efficacy.

Although the examples we considered here involved
low-order basis functions (constant or linear variation
over triangles), it would be straightforward to extend 
the method to higher-order basis functions. Indeed,
switching to basis functions of quadratic or higher 
order would amount simply to choosing different 
$P$ polynomials in (\ref{OriginalIntegral}); the 
computer-algebra methods mentioned in Section 
\ref{ComputerAlgebraSection} could then be used to
identify the corresponding $\mc P$ polynomials
in equations (2) and (5).
A less straightforward but potentially fruitful challenge
would be to extend the method to the case of 
\textit{curved} triangles, in which case the integrand 
of (\ref{OriginalIntegral}) may contain non-polynomial
factors~\cite{Graglia1997}.

Although---as noted in the Introduction---the problem 
of evaluating singular BEM integrals has been 
studied for decades with dozens of algorithms
published, the problem of choosing \textit{which} of 
the myriad available schemes to use in a practical 
BEM solver must surely remain bewildering to 
implementors. A comprehensive comparative 
survey of available methods---including considerations
such as numerical accuracy vs. computation time,
the reuse of previous computations to accelerate
calculations at new frequencies, the availability 
of open-source code implementations, and the complexity
and length of codes versus their extendability and 
flexibility (i.e. the range of possible integrands 
they can handle)---would be an invaluable contribution 
to the literature.

A complete implementation of the method described in 
this paper is incorporated into {\sc scuff-em}, a
free, open-source software implementation of the
boundary-element method~\cite{TaylorDuffyCode}.

\section*{Acknowledgments}

The authors are grateful to A. G. Polimeridis for
many valuable suggestions and for a careful reading
of the manuscript.

This work was supported in part by the Defense Advanced Research
Projects Agency (DARPA) under grant N66001-09-1-2070-DOD, by the Army
Research Office through the Institute for Soldier Nanotechnologies
(ISN) under grant W911NF-07-D-0004, and by the AFOSR Multidisciplinary
Research Program of the University Research Initiative (MURI) for
Complex and Robust On-chip Nanophotonics under grant FA9550-09-1-0704.

\appendices

\section{Expressions for subregion-dependent functions}
\label{SubRegionAppendix}

The TDM formulas (2) refer to functions 
$X_d$ and $\mathcal{P}_{dn}$ indexed by the subregion $d$
of the original 4-dimensional integration domain. In this
Appendix we give detailed expressions for these quantities.
In what follows, the geometric parameters 
$\vb A, \vb B, \vb A^\prime, \vb B^\prime, \vb V_i$ refer to 
Figure \ref{VertexEdgeLabelFigure}, and the 
functions $\vb x(\xi_1,\xi_2)$ and $\vb x^\prime(\eta_1, \eta_2)$
map the standard triangle into $\mc T, \mc T^\prime$
according to
\numeq{VariableTransformation}
{
\vb x(\xi_1, \xi_2) 
 = \vb V_1 + \xi_1 \vb A + \vb \xi_2 \vb B, 
\quad
\vb x^\prime(\eta_1, \eta_2)
 = \vb V_1 + \eta_1 \vb A^\prime + \eta_2 \vb B^\prime.
}
with ranges $0\le \xi_1,\eta_1\le 1$, 
$0\le \xi_2 \le \xi_1$, $0\le \eta_2\le \eta_1.$

\subsection*{Common Triangle}

\paragraph*{$u$ functions} First define ancillary
functions $u_1$ and $u_2$ depending on $y_1$:
\setcounter{equation}{20}
\numeq{CommonTriangleUTable}{
\begin{array}{|c|c|c|} \hline
 d & u_{1d}(y_1) & u_{2d}(y_1) \\ \hline
 1 & 1           & y_1         \\ \hline
 2 & y_1         & (y_1-1)     \\ \hline
 3 & y_1         & 1           \\ \hline
\end{array}
}

\paragraph*{Reduced distance function}
The function 
$X\supt{CT}_d(y_1)$ that enters formulas (2a)
and (5a) is
\begin{align*}
&\!\!X_d\supt{CT}(y_1) 
\\
&= 
  \sqrt{  |\vb A|^2 u_{1d}^2 (y_1) 
         +|\vb B|^2 u_{2d}^2 (y_1) 
         + 2\vb A \cdot \vb B u_{1d} (y_1) u_{2d}(y_1)}.
\end{align*}

\paragraph{$\mathcal{P}$ polynomials}
First define polynomials obtained as definite
integrals over the original $P$ polynomial in
(\ref{OriginalIntegral}):
\begin{align} 
 &\hspace{-0.2in}
 H_d(u_1, u_2) \equiv 
 4AA^\prime
 \int_{\xi_{1d}\sups{lower}}^{\xi_{1d}\sups{upper}} d \xi_1
 \int_{\xi_{2d}\sups{lower}}^{\xi_{2d}\sups{upper}} d \xi_2 
 \,\bigg\{ 
\nn
 &P\Big( \vb x(\xi_1,  \xi_2),
          \vb x^\prime(u_1 + \xi_1, u_2 + \xi_2) 
    \Big)
\nn
 &\,\,+P\Big( \vb x(u_1+ \xi_1,  u_2 + \xi_2),
          \vb x^\prime(\xi_1,\xi_2) 
    \Big) 
 \bigg\}
\label{CommonTriangleHIntegral}
\end{align} 
where $\vb x(\xi_1, \xi_2)$ and $\vb x^\prime(\eta_1, \eta_2)$ 
are as in (\ref{VariableTransformation}) and 
where the limits of the $\xi_1, \xi_2$ integrals are as follows:
\numeq{CommonTriangleIntegrationLimits}
{\begin{array}{|c|c|c|c|c|} \hline
 d & \xi_{1d}\supt{lower} & \xi_{1d}\supt{upper} & 
     \xi_{2d}\supt{lower} & \xi_{2d}\supt{upper}    \\\hline
 1 & 0       & 1-u_1     & 0    & \xi_1             \\ \hline
 2 & -u_2    & 1-u_1     & -u_2 & \xi_1             \\ \hline
 3 & u_2-u_1 & 1-u_1     & 0    & \xi_1 - (u_2-u_1) \\ \hline
\end{array}
}
Now evaluate $H_d$ at $w, y_1$ dependent arguments and
expand the result as a polynomial in $w$ to obtain the 
$\mathcal{P}_{dn}\supt{CT}$ functions:
\numeq{CommonTriangleHToP}
{
   H_d\Big( w u_{1d}(y_1), w u_{2d}(y_1) \Big)
   \equiv \sum_{n} w^n \mathcal{P}_{dn}\supt{CT}(y_1).
}

\subsection*{Common Edge}

\paragraph*{$u,\xi$ functions} First 
define\footnote{This table is similar to 
Table III of \citeasnoun{Taylor2003}, but it 
corrects two errors in that table---namely, 
those in table entries $u_2,E_4$ and $\xi_2,E_1$.}
ancillary functions $u_1, u_2, \xi_2$ depending 
on $y_1$ and $y_2$:
\numeq{CommonEdgeTable}
{
\begin{array}{|c|c|c|c|} \hline
 d & u_{1d}(y_1,y_2) & u_{2d}(y_1,y_2) & \xi_{2d}(y_1, y_2) \\ \hline
 1 & -y_1    &  -y_1y_2      &  1-y_1+y_1 y_2 \\ \hline
 2 &  y_1    &   y_1y_2      &  1-y_1         \\ \hline
 3 & -y_1y_2 &   y_1(1-y_2)  &  1-y_1         \\ \hline
 4 &  y_1y_2 &  -y_1(1-y_2)  &  1-y_1y_2      \\ \hline
 5 & -y_1y_2 &  -y_1         &  1             \\ \hline
 6 &  y_1y_2 &   y_1         &  1-y_1         \\ \hline
\end{array}
}

\paragraph*{Reduced distance function}

The function 
$X\supt{CE}_d(y_1, y_2)$ that enters formulas 
(2b) and (5b) is
\begin{align*}
 &X_d\supt{CE}(y_1, y_2)=
 \bigg[  |\vb A|^2          u_{1d}^2
       +|\vb B^\prime|^2   u_{2d}^2
       +|\vb L|^2          \xi_{2d}^2
\\
 &\,\,\,
       + 2\vb A \cdot \vb B^\prime u_{1d} u_{2d}
       + 2\vb A \cdot \vb L u_{1d} \xi_{2d}
       + 2\vb B^\prime \cdot \vb L u_{2d} \xi_{2d}
 \bigg]^{1/2}
\end{align*}
where $u_{1d}, u_{2d}, \xi_{2d}$ are functions of 
$y_1$ and $y_2$ as in (\ref{CommonEdgeTable}), and 
where $\vb L\equiv \vb B^\prime - \vb B$.

\paragraph*{$\mathcal{P}$ polynomials}

First define polynomials obtained as definite
integrals over the original $P$ polynomial in
(\ref{OriginalIntegral}):
\begin{align*} 
 &\!\! H_d(u_1, u_2, \xi_2 ) 
\\
&\equiv 4AA^\prime
 \int_{\xi_{1d}\sups{lower}}^{\xi_{1d}\sups{upper}} d \xi_1
 \bigg\{ P\Big( \vb x(\xi_1,  \xi_2), 
                \vb x^\prime(u_1 + \xi_1, u_2 + \xi_2) 
          \Big) 
 \bigg\}
\end{align*} 
where $\vb x(\xi_1, \xi_2)$ and $\vb x^\prime(\eta_1, \eta_2)$ 
are as in (\ref{VariableTransformation}) and 
where the limits of the $\xi_1$ integral are as follows:
$$\begin{array}{|c|c|c|c|c|} \hline
 d & \xi_{1d}\supt{lower} & \xi_{1d}\supt{upper} \\ \hline
 1 & \xi_2 + (u_2-u_1) & 1     \\ \hline
 2 & \xi_2             & 1-u_1 \\ \hline
 3 & \xi_2 + (u_2-u_1) & 1     \\ \hline
 4 & \xi_2             & 1-u_1 \\ \hline
 5 & \xi_2             & 1     \\ \hline
 6 & \xi_2 + (u_2-u_1) & 1-u_1 \\ \hline
\end{array}.
$$
Now evaluate $H_d$ at $w, y_1, y_2$-dependent arguments and
expand the result as a polynomial in $w$ to obtain the 
$\mathcal{P}_{dn}\supt{CE}$ functions:
\begin{align}
&H_d\Big( w u_{1d}(y_1,y_2), w u_{2d}(y_1, y_2) ,
            w \xi_2(y_1,y_2) \Big)
\nn
&\quad
   \equiv \sum_{n} w^n \mathcal{P}_{dn}\supt{CE}(y_1, y_2).
\label{CommonEdgeHToP}
\end{align}

\subsection*{Common Vertex}

\paragraph*{$\xi,\eta$ functions} First define ancillary
functions $\xi_1, \xi_2, \eta_1, \eta_2$ 
depending on $y_1,y_2,y_3$:

\numeq{CommonVertexTable}
{
\begin{array}{|c|c|c|c|c|} \hline
 d & \xi_{1d}(\vb y) & \xi_{2d}(\vb y) & \eta_{1d}(\vb y) & \eta_{2d}(\vb y) 
\\\hline
 1 & 1 & y_1     & y_2 & y_2 y_3 
\\\hline
 2 & y_2 & y_2 y_3 & 1   & y_1 
\\\hline
\end{array}
}

\paragraph*{Reduced distance function} The function 
$X\supt{CV}_d(y_1, y_2, y_3)$ that enters formulas 
(2c) and (5c) is 
\begin{align*}
&\hspace{-0.1in} X_d\supt{CV}(y_1, y_2, y_3)
\\
= \bigg[ & |\vb A|^2           \xi_{1d}^2
          +|\vb B|^2           \xi_{2d}^2
          +|\vb A^\prime|^2    \eta_{1d}^2
          +|\vb B^\prime|^2    \eta_{1d}^2
\\      
   &     + 2\big(\vb A \cdot \vb B\big)        \xi_{1d} \xi_{2d}
         - 2\big(\vb A \cdot \vb A^\prime\big) \xi_{1d} \eta_{1d}
\\[5pt]
   &
         - 2\big(\vb A \cdot \vb B^\prime\big) \xi_{1d} \eta_{2d}
         - 2\big(\vb B \cdot \vb A^\prime\big) \xi_{2d} \eta_{1d}
\\
   &
         - 2\big(\vb B \cdot \vb B^\prime\big) \xi_{2d} \eta_{2d}
         + 2\big(\vb A^\prime \cdot \vb B^\prime\big) \eta_{1d} \eta_{2d}
  \bigg]^{1/2}
\end{align*}
where $\xi$ and $\eta$ are functions of
$y_1, y_2,$ and $y_3$ as in (\ref{CommonVertexTable}).

\paragraph*{$\mathcal{P}$ polynomials}

In contrast to the common-triangle and common-edge cases,
in the common-vertex case there is no integration over the 
original $P$ polynomial; instead, we simply evaluate the original
$P$ polynomial at $w$- and $\vb y$-dependent arguments,
expand the result as a power series in $w$, and identify
the coefficients of this power series as the 
$\mathcal{P}(\vb y)$ polynomials:

\numeq{CommonVertexHToP}
{P\Big( w \xi_{1d}(\vb y), w \xi_{2d}(\vb y), 
          w \eta_{1d}(\vb y), w \eta_{2d}(\vb y)
    \Big)
   \equiv \sum_{n} w^n \mathcal{P}_{dn}\supt{CV}(\vb y).
}

\subsection*{The $\alpha$, $\beta$, $\gamma$ Coefficients}

For twice-integrable kernels, the master TDM formulas 
(5) refer
to parameters $\alpha$, $\beta$, and $\gamma$ defined 
for the various cases and the various subregions. These
parameters are defined by completing the square under the
radical sign in the reduced-distance functions $X_d$:

\begin{align*}
 X_d\supt{CT}(y_1) 
  &\equiv
   \sqrt{ \alpha\supt{CT}_d 
          (y_1 + \beta\supt{CT}_d)^2 
         + \gamma\supt{CT}_d }
\\
 X_d\supt{CE}(y_1, y_2) 
  &\equiv
   \sqrt{ \alpha\supt{CE}_d 
          (y_2 + \beta\supt{CE}_d)^2 
         + \gamma\supt{CE}_d }
\\
 X_d\supt{CV}(y_1, y_2, y_3) 
  &\equiv
   \sqrt{ \alpha\supt{CV}_d 
          (y_3 + \beta\supt{CV}_d)^2 
         + \gamma\supt{CV}_d }
\end{align*}
In all cases, the $\{\alpha, \beta, \gamma\}$ coefficients depend 
on the geometric parameters ($\vb A, \vb B,$ etc.). In the 
CE case, they depend additionally on the variable $y_1$, and in
the CV case they depend additionally 
on the variables $y_1$ and $y_2$.

\section{First and Second Kernel Integrals}
\label{FirstSecondIntegralAppendix}

In this Appendix we collect expressions for the first and 
second integrals of various commonly encountered kernel
functions.

\subsection*{First Kernel Integrals}

In Section \ref{GeneralizedTaylorDuffySection} we 
defined the ``first integral'' of a kernel function 
$K(r)$ to be

$$ \mathcal{K}_n(X) \equiv \int_0^1 w^n K(wX) \, dw.$$

\noindent First integrals for some commonly encountered kernels
are presented in Table \ref{FirstIntegralTable}.

\begin{table}[h]
\begin{center}
\renewcommand{\arraystretch}{2.5}
$$\begin{array}{|c|c|} \hline
 \ K(r)
  & 
  \mathcal{K}_n(X)
\\\hline
  r^p 
  & 
  \displaystyle{ \frac{X^p}{1+n+p} } \vphantom{\int_\int^\int}
\\\hline
  \displaystyle{ \frac{e^{ikr}}{4\pi r} }
  & 
  \displaystyle{ \frac{e^{ikX}}{4\pi n X} \text{ExpRel}(n,-ikX)
               }
\\[5pt]\hline
  \displaystyle{ (ikr-1)\frac{e^{ikr}}{4\pi r^3} }
  &
  \parbox{0.3\textwidth}
   { \begin{align*}
       &\displaystyle{ \frac{e^{ikX}}{4\pi X}
                      \bigg[ \frac{ik}{(n-1)X}\text{ExpRel}(n-1,-ikX) 
                    }\\
       &\qquad\qquad
        \displaystyle{ -\frac{1}{(n-2)X^2}\text{ExpRel}(n-2,-ikX)
                        \bigg]
                     }
     \end{align*}
   } 
  \vphantom{\int_\int^\int}
\\\hline
\end{array}$$
\end{center}
\caption{First integrals for some relevant kernel functions.}
\label{FirstIntegralTable}
\end{table}
\renewcommand{\arraystretch}{1.0}

The function \text{ExpRel}$(n,z)$ in Table 
\ref{FirstIntegralTable} is
the $n$th ``relative exponential'' 
function,\footnote{Our terminology here is borrowed from
the {\sc gnu} Scientific Library,
\texttt{http://www.gnu.org/software/gsl/}.}
defined as the usual exponential function with 
the first $n$ terms of its power-series expansion 
subtracted and the result normalized to have 
value 1 at $z=0$:
\begin{subequations}
\begin{align}
 \text{ExpRel}(n,z) 
&\equiv \frac{n!}{z^n}\Big[ e^{z} - 1 - z - \frac{z^2}{2} 
                       - \cdots - \frac{z^{n-1}}{(n-1)!}
                 \Big]
\label{ExpRelA}
\\
&= 1 + \frac{z}{(n+1)} + \frac{z^2}{(n+1)(n+2)} + \cdots
\label{ExpRelB}
\end{align}
\end{subequations}
For $|z|$ small,
the relative exponential function may be 
computed using the rapidly convergent series expansion 
(\ref{ExpRelB}), and indeed for computational purposes
at small $z$ it is important \textit{not} to use the defining 
expression (\ref{ExpRelA}), na\"ive use of which
invites catastrophic loss of numerical precision. For
example, at $z=10^{-4}$, each term subtracted from $e^z$ 
in the square brackets in (\ref{ExpRelA}) eliminates 4 digits
of precision, so that in standard 64-bit floating-point arithmetic 
a calculation of $\text{ExpRel}(3,z)$ would be accurate only to 
approximately 3 digits, while a calculation of $\text{ExpRel}(4,z)$ 
would yield pure numerical noise.

On the other hand, for values of $z$ with large negative real
part---which arise in calculations involving lossy materials
at short wavelengths---it is most convenient to evaluate 
\text{ExpRel} in a different way, as discussed below.


\subsection*{Second Kernel Integrals for the $r^p$ kernel}

In Section \ref{TwiceIntegrableKernelSection} we defined the
``second integrals'' of a kernel function $K(r)$ to be the 
following definite integrals involving the first integral:
\begin{align}
  \mathcal{J}_n(\alpha,\beta,\gamma) 
&\equiv 
  \int_0^1 \mathcal{K}_n\Big( \alpha\sqrt{ (y+\beta)^2 + \gamma^2} \Big) \, dy
\\
  \mathcal{L}_n(\alpha,\beta,\gamma) 
&\equiv 
  \int_0^1 y\, \mathcal{K}_n\Big( \alpha\sqrt{ (y+\beta)^2 + \gamma^2} \Big) \, dy.
\label{SecondIntegralsAgain}
\end{align}
\noindent For the particular kernel function $K(r)=r^p$ with
(positive or negative) integer power $p$, the second integrals
are the following analytically-evaluatable integrals:
\begin{align}
 \mathcal{J}_n(\alpha,\beta,\gamma)
&=\frac{\alpha^p}{(1+p+n)}
  \underbrace{ \int_0^1 \Big[(y+\beta)^2 + \gamma^2]^{p/2} \, dy
             }_{\equiv \overline{\mc J_p}(\beta,\gamma)}
\\
 \mathcal{L}_n(\alpha,\beta,\gamma)
&=\frac{\alpha^p}{(1+p+n)}
   \underbrace{\int_0^1 y \Big[(y+\beta)^2 + \gamma^2]^{p/2} \, dy
              }_{\equiv \overline{\mc L_p}(\beta,\gamma)}
\end{align}
The integral arising in the first line here is tabulated below
for a few values of $p$. (The table is easily extended to 
arbitrary positive or negative values of $p$.)
In this table, we have 
$S=\sqrt{\beta^2 + \gamma^2}, T=\sqrt{(\beta+1)^2+\gamma^2}.$

$$
 \renewcommand{\arraystretch}{2.5}
 \begin{array}{|c|c|}\hline
     p & \overline{\mc J_p} 
         \equiv \int_0^1 \Big[(y+\beta)^2 + \gamma^2\Big]^{p/2} \, dy 
  \\ \hline
     -3
   & \frac{1}{\gamma^2}\Big[ \frac{\beta+1}{T} - \frac{\beta}{S}
                       \Big]
\\ \hline
     -2
   & \frac{1}{\gamma}\Big[ \arctan\frac{\beta+1}{\gamma} - 
                           \arctan\frac{\beta}{\gamma} 
                     \Big]
  \\ \hline
     -1
   & \ln \frac{\beta+1 + T }{\beta + S} 
  \\ \hline
      1
   &  \frac{1}{2}\Big[ \beta(T-S) + T 
                       + \gamma^2 \overline{\mc J_{-1}}
                 \Big]
  \\ \hline
      2
   &  \frac{1}{2}\Big[ T^2  + S^2 - \frac{1}{2}\Big]
  \\ \hline
 \end{array}
 \renewcommand{\arraystretch}{1.0}
$$

The $\overline{\mc L}$ functions are related to the $\overline{\mc J}$ 
functions according to 
$$
\overline{\mc L_p} = 
 \begin{cases}
   \displaystyle{
      -\beta \overline{\mc J_p} 
    + \frac{1}{p+2}\left(T^{p+2} - S^{p+2}\right),
                }  \qquad &p\ne -2 
   \\[10pt]
  \displaystyle{
      -\beta \overline{\mc J_p} 
    + \ln \left(\frac{T}{S}\right),
               }  \qquad &p=-2.
  \end{cases}
$$


\subsection*{Second kernel integrals for the
             Helmholtz kernel in the short-wavelength limit}

For the EFIE kernel $K(r)=e^{ikr}/(4\pi r)$, 
the first integral $\mathcal{K}_n(X)$ (Table \ref{FirstIntegralTable})
involves the quantity $e^{ikX}\text{ExpRel}(n,-ikX).$ As noted
above, for small values of $|kX|$ the relative exponential
function is well represented by the first few terms in the 
expansion (\ref{ExpRelB}).
However, for $|kX|$ large it is convenient instead to
use the defining expression (\ref{ExpRelA}), in terms of 
which we find
\begin{align}
& e^{ikX}\text{ExpRel}(n,-ikX)
\label{ExpRelLargeK}\\
&\hspace{0.1in}=\frac{n!}{(-ikX)^n} 
\nn
&\qquad-
  e^{ikX}\bigg[ \frac{n^!}{(-ikX)^n}
  \Big(1 - ikX + \cdots + \frac{(-ikX)^{n-1}}{(n-1)!}
  \Big)\bigg].
\nonumber
\end{align}
For $k$ values with large positive imaginary part,
the first term here decays algebraically with $|k|$,
while the second term decays \textit{exponentially} and
hence makes negligible contribution to the sum when
$|k|$ is sufficiently large.
This suggests that, for $k$ values with large positive imaginary
part, we may approximate the first kernel integrals 
in Table \ref{FirstIntegralTable} by retaining only the 
first term in (\ref{ExpRelLargeK}). This leads to the
following approximate expressions for the first kernel
integrals in Table (\ref{FirstIntegralTable}):
$$
 \begin{array}{l}
   \displaystyle{K(r)=\frac{e^{ikr}}{4\pi r}}
\\[6pt]
   \quad\quad \Longrightarrow \,\,
   \mc K_n(X) \,\,\xrightarrow{\text{Im k}\to\infty}\,\,
   \displaystyle{\frac{(n-1)!}{4\pi(-ik)^n X^{n+1}}}
\\[12pt]
   \displaystyle{K(r)=(ikr-1)\frac{e^{ikr}}{4\pi r^3}}
\\[6pt]
   \quad\quad\Longrightarrow \,\,
   \mc K_n(X) \,\,\xrightarrow{\text{Im k}\to\infty}\,\,
   \displaystyle{-\frac{(n-1)[(n-3)!]}{4\pi(-ik)^{n-2} X^{n+1}}}
 \end{array}
$$
The important point here is that the simpler $X$ dependence
of these $\mathcal{K}$ functions 
in the $\text{Im }k\to\infty$ limit renders these kernels
\textit{twice integrable}. This allows us to make use of the 
twice-integrable versions of the TDM formulas to compute integrals 
involving these kernels in this limit. In particular, we find the 
following second integrals:

\subsection*{For $K(r)=\frac{e^{ikr}}{4\pi r}$ as $\text{Im }k\to \infty$:}


\begin{align*}
\mathcal{J}_n(\alpha,\beta,\gamma) 
 &\to \frac{(n-1)!}{4\pi (-ik)^n \alpha^{n+1}}
   \,
    \overline{\mc J}_{-(n+1)/2}(\beta,\gamma)
\\
\mathcal{L}_n(\alpha,\beta,\gamma) 
 &\to \frac{(n-1)!}{4\pi (-ik)^n \alpha^{n+1}}
   \,
    \overline{\mc L}_{-(n+1)/2}(\beta,\gamma)
\end{align*}

\subsection*{For $K(r)=(ikr-1)\frac{e^{ikr}}{4\pi r^3}$ as $\text{Im }k\to \infty$:}

\begin{align*}
\mathcal{J}_n(\alpha,\beta,\gamma) 
 &\to -\frac{(n-1)[(n-3)!]}{4\pi (-ik)^{n-2} \alpha^{n+1}}
   \,
    \overline{\mc J}_{-(n+1)/2}(\beta,\gamma)
\\
\mathcal{L}_n(\alpha,\beta,\gamma) 
 &\to -\frac{(n-1)[(n-3)!]}{4\pi (-ik)^{n-2} \alpha^{n+1}}
   \,
    \overline{\mc L}_{-(n+1)/2}(\beta,\gamma)
\\
\end{align*}

The $\overline{\mc J}, \overline{\mc L}$ functions
were evaluated above in the discussion of the 
$K(r)=r^p$ kernel.

\bibliographystyle{IEEEtran}

\begin{thebibliography}{10}
\providecommand{\url}[1]{#1}
\csname url@samestyle\endcsname
\providecommand{\newblock}{\relax}
\providecommand{\bibinfo}[2]{#2}
\providecommand{\BIBentrySTDinterwordspacing}{\spaceskip=0pt\relax}
\providecommand{\BIBentryALTinterwordstretchfactor}{4}
\providecommand{\BIBentryALTinterwordspacing}{\spaceskip=\fontdimen2\font plus
\BIBentryALTinterwordstretchfactor\fontdimen3\font minus
  \fontdimen4\font\relax}
\providecommand{\BIBforeignlanguage}[2]{{%
\expandafter\ifx\csname l@#1\endcsname\relax
\typeout{** WARNING: IEEEtran.bst: No hyphenation pattern has been}%
\typeout{** loaded for the language `#1'. Using the pattern for}%
\typeout{** the default language instead.}%
\else
\language=\csname l@#1\endcsname
\fi
#2}}
\providecommand{\BIBdecl}{\relax}
\BIBdecl

\bibitem{TaylorDuffyCode}
\texttt{http://homerreid.com/scuff-EM/SingularIntegrals}.

\bibitem{Taylor2003}
D.~Taylor, ``Accurate and efficient numerical integration of weakly singular
  integrals in {G}alerkin {EFIE} solutions,'' \emph{Antennas and Propagation,
  IEEE Transactions on}, vol.~51, no.~7, pp. 1630--1637, 2003.

\bibitem{Duffy1982}
M.~G. Duffy, ``Quadrature over a pyramid or cube of integrands with a
  singularity at a vertex,'' \emph{SIAM Journal on Numerical Analysis},
  vol.~19, no.~6, pp. 1260--1262, 1982.

\bibitem{Taskinen2003}
P.~Yla-Oijala and M.~Taskinen, ``Calculation of {CFIE} impedance matrix
  elements with {RWG} and nx{RWG} functions,'' \emph{Antennas and Propagation,
  IEEE Transactions on}, vol.~51, no.~8, pp. 1837--1846, 2003.

\bibitem{Jarvenpaa2006}
S.~Jarvenpaa, M.~Taskinen, and P.~Yla-Oijala, ``Singularity subtraction
  technique for high-order polynomial vector basis functions on planar
  triangles,'' \emph{Antennas and Propagation, IEEE Transactions on}, vol.~54,
  no.~1, pp. 42--49, 2006.

\bibitem{TongChew2007}
\BIBentryALTinterwordspacing
M.~S. Tong and W.~C. Chew, ``Super-hyper singularity treatment for solving 3d
  electric field integral equations,'' \emph{Microwave and Optical Technology
  Letters}, vol.~49, no.~6, pp. 1383--1388, 2007. [Online]. Available:
  \url{http://dx.doi.org/10.1002/mop.22443}
\BIBentrySTDinterwordspacing

\bibitem{Klees1996}
\BIBentryALTinterwordspacing
R.~Klees, ``\BIBforeignlanguage{English}{Numerical calculation of weakly
  singular surface integrals},'' \emph{\BIBforeignlanguage{English}{Journal of
  Geodesy}}, vol.~70, no.~11, pp. 781--797, 1996. [Online]. Available:
  \url{http://dx.doi.org/10.1007/BF00867156}
\BIBentrySTDinterwordspacing

\bibitem{Cai2002}
\BIBentryALTinterwordspacing
W.~Cai, Y.~Yu, and X.~C. Yuan, ``Singularity treatment and high-order {RWG}
  basis functions for integral equations of electromagnetic scattering,''
  \emph{International Journal for Numerical Methods in Engineering}, vol.~53,
  no.~1, pp. 31--47, 2002. [Online]. Available:
  \url{http://dx.doi.org/10.1002/nme.390}
\BIBentrySTDinterwordspacing

\bibitem{Khayat2005}
M.~Khayat and D.~Wilton, ``Numerical evaluation of singular and near-singular
  potential integrals,'' \emph{Antennas and Propagation, IEEE Transactions on},
  vol.~53, no.~10, pp. 3180--3190, 2005.

\bibitem{Ismatullah2008}
I.~Ismatullah and T.~Eibert, ``Adaptive singularity cancellation for efficient
  treatment of near-singular and near-hypersingular integrals in surface
  integral equation formulations,'' \emph{Antennas and Propagation, IEEE
  Transactions on}, vol.~56, no.~1, pp. 274--278, 2008.

\bibitem{Graglia2008}
R.~Graglia and G.~Lombardi, ``Machine precision evaluation of singular and
  nearly singular potential integrals by use of {G}auss quadrature formulas for
  rational functions,'' \emph{Antennas and Propagation, IEEE Transactions on},
  vol.~56, no.~4, pp. 981--998, 2008.

\bibitem{Polimeridis2010}
\BIBentryALTinterwordspacing
A.~G. Polimeridis and J.~R. Mosig, ``Complete semi-analytical treatment of
  weakly singular integrals on planar triangles via the direct evaluation
  method,'' \emph{International Journal for Numerical Methods in Engineering},
  vol.~83, no.~12, pp. 1625--1650, 2010. [Online]. Available:
  \url{http://dx.doi.org/10.1002/nme.2877}
\BIBentrySTDinterwordspacing

\bibitem{Polimeridis2013}
A.~Polimeridis, F.~Vipiana, J.~Mosig, and D.~Wilton, ``{DIRECTFN}: Fully
  numerical algorithms for high precision computation of singular integrals in
  {G}alerkin {SIE} methods,'' \emph{Antennas and Propagation, IEEE Transactions
  on}, vol.~61, no.~6, pp. 3112--3122, 2013.

\bibitem{Andra1997}
H.~Andrä and E.~Schnack, ``\BIBforeignlanguage{English}{Integration of
  singular {G}alerkin-type boundary element integrals for 3d elasticity
  problems},'' \emph{\BIBforeignlanguage{English}{Numerische Mathematik}},
  vol.~76, no.~2, pp. 143--165, 1997.

\bibitem{SauterSchwab2010}
\BIBentryALTinterwordspacing
S.~Sauter and C.~Schwab, \emph{Boundary Element Methods}, ser. Springer series
  in computational mathematics.\hskip 1em plus 0.5em minus 0.4em\relax
  Springer, 2010. [Online]. Available:
  \url{http://books.google.com/books?id=yEFu7sVW3LEC}
\BIBentrySTDinterwordspacing

\bibitem{Erichsen1998}
\BIBentryALTinterwordspacing
S.~Erichsen and S.~A. Sauter, ``Efficient automatic quadrature in 3-d
  {G}alerkin {BEM},'' \emph{Computer Methods in Applied Mechanics and
  Engineering}, vol. 157, no. 3–4, pp. 215 -- 224, 1998, <ce:title>Papers
  presented at the Seventh Conference on Numerical Methods and Computational
  Mechanics in Science and Engineering</ce:title>. [Online]. Available:
  \url{http://www.sciencedirect.com/science/article/pii/S0045782597002363}
\BIBentrySTDinterwordspacing

\bibitem{Harrington93}
R.~F. Harrington, \emph{Field Computation by Moment Methods}.\hskip 1em plus
  0.5em minus 0.4em\relax Wiley-IEEE Press, 1993.

\bibitem{Chew2009}
\BIBentryALTinterwordspacing
W.~Chew, M.~Tong, and B.~Hu, \emph{Integral Equation Methods for
  Electromagnetic and Elastic Waves}, ser. Synthesis Lectures on Computational
  Electromagnetics Series.\hskip 1em plus 0.5em minus 0.4em\relax Morgan \&
  Claypool Publishers, 2009. [Online]. Available:
  \url{http://books.google.com/books?id=PJN9meadzT8C}
\BIBentrySTDinterwordspacing

\bibitem{Botha2013}
M.~Botha, ``A family of augmented {D}uffy transformations for near-singularity
  cancellation quadrature,'' \emph{Antennas and Propagation, IEEE Transactions
  on}, vol.~61, no.~6, pp. 3123--3134, June 2013.

\bibitem{Vipiana2013}
F.~Vipiana, D.~Wilton, and W.~Johnson, ``Advanced numerical schemes for the
  accurate evaluation of 4-d reaction integrals in the method of moments,''
  \emph{Antennas and Propagation, IEEE Transactions on}, vol.~PP, no.~99, pp.
  1--1, 2013.

\bibitem{Caorsi1993}
P.~Caorsi, D.~Moreno, and F.~Sidoti, ``Theoretical and numerical treatment of
  surface integrals involving the free-space {G}reen's function,''
  \emph{Antennas and Propagation, IEEE Transactions on}, vol.~41, no.~9, pp.
  1296--1301, 1993.

\bibitem{Eibert1995}
T.~Eibert and V.~Hansen, ``On the calculation of potential integrals for linear
  source distributions on triangular domains,'' \emph{Antennas and Propagation,
  IEEE Transactions on}, vol.~43, no.~12, pp. 1499--1502, 1995.

\bibitem{Sarkar1984}
T.~Sarkar and R.~Harrington, ``The electrostatic field of conducting bodies in
  multiple dielectric media,'' \emph{Microwave Theory and Techniques, IEEE
  Transactions on}, vol.~32, no.~11, pp. 1441--1448, 1984.

\bibitem{RWG1982}
S.~Rao, D.~Wilton, and A.~Glisson, ``Electromagnetic scattering by surfaces of
  arbitrary shape,'' \emph{Antennas and Propagation, IEEE Transactions on},
  vol.~30, no.~3, pp. 409--418, May 1982.

\bibitem{Rius2001}
J.~Rius, E.~Ubeda, and J.~Parron, ``On the testing of the magnetic field
  integral equation with {R}{W}{G} basis functions in method of moments,''
  \emph{Antennas and Propagation, IEEE Transactions on}, vol.~49, no.~11, pp.
  1550--1553, Nov 2001.

\bibitem{Medgyesi1994}
\BIBentryALTinterwordspacing
L.~N. Medgyesi-Mitschang, J.~M. Putnam, and M.~B. Gedera, ``Generalized method
  of moments for three-dimensional penetrable scatterers,'' \emph{J. Opt. Soc.
  Am. A}, vol.~11, no.~4, pp. 1383--1398, Apr 1994. [Online]. Available:
  \url{http://josaa.osa.org/abstract.cfm?URI=josaa-11-4-1383}
\BIBentrySTDinterwordspacing

\bibitem{Taskinen2005}
P.~Yla-Oijala and M.~Taskinen, ``Well-conditioned {M}\"uller formulation for
  electromagnetic scattering by dielectric objects,'' \emph{Antennas and
  Propagation, IEEE Transactions on}, vol.~53, no.~10, pp. 3316--3323, 2005.

\bibitem{libSGJC}
\texttt{http://ab-initio.mit.edu/wiki/index.php/Cubature}.

\bibitem{Khayat2008}
M.~Khayat, D.~Wilton, and P.~Fink, ``An improved transformation and optimized
  sampling scheme for the numerical evaluation of singular and near-singular
  potentials,'' \emph{Antennas and Wireless Propagation Letters, IEEE}, vol.~7,
  pp. 377--380, 2008.

\bibitem{Cools2003}
\BIBentryALTinterwordspacing
R.~Cools, ``An encyclopaedia of cubature formulas,'' \emph{Journal of
  Complexity}, vol.~19, no.~3, pp. 445 -- 453, 2003, oberwolfach Special Issue.
  [Online]. Available:
  \url{http://www.sciencedirect.com/science/article/pii/S0885064X03000116}
\BIBentrySTDinterwordspacing

\bibitem{Graglia1997}
R.~Graglia, D.~Wilton, and A.~Peterson, ``Higher order interpolatory vector
  bases for computational electromagnetics,'' \emph{Antennas and Propagation,
  IEEE Transactions on}, vol.~45, no.~3, pp. 329--342, 1997.

\end{thebibliography}

\end{document}